\documentclass{aa}
\usepackage{amsmath}
\usepackage{cases}
\usepackage{txfonts}
\usepackage{graphicx}
\usepackage{natbib}
\usepackage{epstopdf}
\usepackage{nicefrac}
\usepackage{hyperref}
\usepackage{threeparttable}
\usepackage[dvipsnames]{xcolor}
\usepackage{longtable,times,multirow}
\usepackage{threeparttable}

\begin{document}

\title{J-PLUS: Stellar Parameters, C, N, Mg, Ca and [$\alpha$/Fe] Abundances for Two Million Stars from DR1}
\subtitle{}
\author{Lin Yang\inst{1}
\and Haibo Yuan\inst{2} 
\and Maosheng Xiang\inst{3} 
\and Fuqing Duan\inst{1}
\and Yang Huang\inst{4}
\and Jifeng Liu\inst{5,6}
\and Timothy C. Beers\inst{7}
\and Carlos Andrés Galarza\inst{8}
\and Simone Daflon\inst{8}
\and J.A. Fern\'andez-Ontiveros\inst{9}
\and Javier Cenarro\inst{9}
\and David Cristóbal-Hornillos\inst{9}
\and Carlos Hernández-Monteagudo\inst{10,11} 
\and Carlos López-Sanjuan\inst{9} 
\and Antonio Marín-Franch\inst{9}
\and Mariano Moles\inst{9} 
\and Jesús Varela\inst{9} 
\and Héctor Vázquez Ramió\inst{9} 
\and Jailson Alcaniz\inst{8}
\and Renato Dupke\inst{8}
\and Alessandro Ederoclite\inst{12}
\and Laerte Sodré Jr.\inst{12}
\and Raul E. Angulo\inst{13,14}
}

\institute{College of Artificial Intelligence, Beijing Normal University No.19, Xinjiekouwai St, Haidian District, Beijing, 100875, P.R.China\\ \email{fqduan@bnu.edu.cn} 
\and
Department of Astronomy, Beijing Normal University No.19, Xinjiekouwai St, Haidian District, Beijing, 100875, P.R.China\\
\email{yuanhb@bnu.edu.cn}
\and
Max-Planck Institute for Astronomy, K\"onigstuhl 17, D-69117 Heidelberg, Germany
\and
South-Western Institute for Astronomy Research, Yunnan University, Kunming 650500, P. R. China
\and
Key Laboratory of Optical Astronomy, National Astronomical Observatories, Chinese Academy of Sciences, Beijing 100012, P. R. China
\and
University of Chinese Academy of Sciences, Beijing 100049, P. R. China
\and
Department of Physics and JINA Center for the Evolution of the Elements
(JINA-CEE), University of Notre Dame, Notre Dame, IN 46556, USA
\and
Observat\'orio Nacional, Rua General Jos\'e Cristino, 77 - Bairro Imperial de S\~ao Crist\'ov\~ao, 20921-400 Rio de Janeiro, Brazil
\and
Centro de Estudios de F\'{\i}sica del Cosmos de Arag\'on (CEFCA), Unidad Asociada al CSIC, Plaza San Juan 1, 44001 Teruel, Spain
\and
Instituto de Astrofísica de Canarias, Calle Vía Láctea SN, 
ES38205 La Laguna, Spain
\and
Departamento de Astrofísica, Universidad de La Laguna, 
ES38205, La Laguna, Spain
\and
Instituto de Astronomia, Geof\'{\i}sica e Ci\^encias Atmosf\'ericas, Universidade de S\~ao Paulo, 05508-090 S\~ao Paulo, Brazil
\and
Donostia International Physics Centre (DIPC), Paseo Manuel de Lardizabal 4, 20018 Donostia-San Sebastián, Spain
\and
Ikerbasque, Basque Foundation for Science, E-48013 Bilbao, Spain
}

\date{Received  2021 / Accepted  2021}

\abstract{The Javalambre Photometric Local Universe Survey (J-PLUS) has obtained precise photometry in twelve specially designed filters for large numbers of Galactic stars. Deriving their precise stellar atmospheric parameters and individual elemental abundances is crucial for studies of Galactic structure, and the assembly history and chemical evolution of our Galaxy.}
{Our goal is to estimate not only stellar parameters (effective temperature, $T_{\rm eff}$, surface gravity, $\log g$, and metallicity, [Fe/H]), but also [$\alpha$/Fe] and four elemental abundances ([C/Fe], [N/Fe], [Mg/Fe], and [Ca/Fe]) using data from J-PLUS DR1.}
{By combining recalibrated photometric data from J-PLUS DR1, {\it Gaia} DR2, and spectroscopic labels from LAMOST, we design and train a set of cost-sensitive neural networks, the {\it CSNet}, to learn the non-linear mapping from stellar colors to their labels. 
Special attention is paid to the poorly populated regions of the label space by giving different weights according to their density distribution. }
{We have achieved precisions of $\delta\, T_{\rm eff}\sim$ 55\,K, $\delta \log g\sim 0.15$\,dex, and $\delta$ [Fe/H] $\sim$ 0.07\,dex, respectively, over a wide range of temperature, surface gravity, and metallicity. The uncertainties of the abundance estimates for [$\alpha$/Fe] and the four individual elements are in the range 0.04--0.08\,dex. We compare our parameter and abundance estimates with those from other spectroscopic catalogs such as APOGEE and GALAH, and find an overall good agreement.}
{Our results demonstrate the potential of well-designed, high-quality photometric data for
determinations of stellar parameters as well as individual elemental abundances.
Applying the method to J-PLUS DR1, we have obtained the aforementioned parameters for about two million stars, providing an outstanding data set for chemo-dynamic analyses of the Milky Way. 
The catalog of the estimated parameters is publicly accessible. 
}
 

\titlerunning{J-PLUS DR1 stellar labels with {\it CSNet}}
 
\maketitle

\section{Introduction} \label{introduction}
Precise determinations of basic stellar parameters and elemental abundances (hereafter referred to as stellar labels) play a fundamental role in a number of fields, including stellar physics, Galactic structure,  the formation and chemical evolution of the Galaxy, and the distribution and properties of dust in the Galaxy. Stellar labels can be determined both spectroscopically and photometrically. These approaches are broadly complementary, each having advantages and disadvantages.   

Currently, a number of large-scale photometric surveys, e.g., the SkyMapper Southern Survey (SMSS):  DR1.1 -- \citealt{Wolf2018}, DR2 -- \citealt{Onken2019}, the Stellar Abundance and Galactic Evolution (SAGE): \citealt{Zheng2018}, the Javalambre Physics of the accelerating universe Astrophysical Survey (J-PAS): \citealt{Benitez2014}, J-PLUS: \citealt{Cenarro2019}, and the Southern Photometric Local Universe Survey (S-PLUS): \citealt{Mendes2019} are producing huge amounts of valuable photometric data for tens of millions of astronomical objects. The medium- and narrow-band filters of these photometric surveys are designed and optimized for precision measurements of key stellar features, opening up a new era of precise and accurate stellar label determinations \citep[see, e.g.,][]{Bailer2002,Arnadottir2010}. 

A number of different empirical and theoretical approaches have been developed to determine stellar labels from the use of photometric data. Thanks to the modern large-scale spectroscopic surveys, such as 
SDSS/SEGUE (\citealt{Yanny2009}), LAMOST (\citealt{Deng2012,Liu2014}), and SDSS/APOGEE (\citealt{Majewski2017}), and their precise estimates of stellar parameters, advanced (high-quality, large sample size, and good coverage in parameter space) training and calibration data sets are available for inferring stellar labels from photometry. These approaches are now capable of delivering photometric stellar labels with comparable precision to spectroscopy for high-quality photometry. 
Using a tool based on empirical metallicity-dependent stellar loci, \citet{Yuan2015b,Yuan2015c} estimated photometric metallicites for a half million FGK dwarf stars in SDSS/Stripe 82,
with a typical error of $\delta$ [Fe/H] $\sim$ 0.1 to 0.2\,dex. Later, \citet{Zhang2021} obtained
metallicity-dependent stellar loci for red-giant stars, which were then used to derive metallicities of 
giants to a precision of $\delta$ [Fe/H] $\sim$ 0.20 to 0.25\,dex, and discriminate metal-poor red giants from main-sequence stars based on SDSS photometry. 

From corrected broad-band {\it Gaia} EDR3 colors alone (\citealt{Niu2021b}; \citealt{Yang2021}), \citet{Xu2021} have determined reliable metallicity estimates for a magnitude-limited sample of about 27 million stars down to [Fe/H] = $-2.5$. Considering that the specially designed SkyMapper
filters {\it uvgriz} of the SMSS are more sensitive to stellar atmospheric parameters than the SDSS filters, \citet{Huang2019} used 
different polynomials to build empirical relations between atmospheric parameters and photometric colors for 
red-giant stars, and derived accurate atmospheric parameters (e.g., effective temperature, $T_{\rm eff}$, surface gravity, $\log g$, and metallicity, [Fe/H]) for about one million red-giant stars from SMSS DR1.1.
\citet{Chiti2020} describe a grid-based synthetic photometry approach, which was employed by \citet{Chiti2021} to obtain photometric metallicities for over a quarter million giants from SMSS DR2. 
With the re-calibrated SMSS DR2 and {\it Gaia} EDR3, \citet{Huang2021a} have further determined
metallicities for over some 24 million stars with a technique similar to the metallicity-dependent stellar loci. 
Thanks to the strong metallicity sensitivity of the SMSS $uv$ filters, the correction of systematic calibration errors in SMSS DR2, and 
the use of well-selected training datasets, the achieved precision is comparable to or slightly better than that derived from spectroscopy, for stars with metallicity as low as [Fe/H] $\sim -3.5$.

In addition to empirical metallicity-dependent stellar loci, machine learning methods such as the random forest algorithm \citep{Miller2015, Bai2019,Andres2021}, Bayesian inference \citep{Bailer2011}, and artificial neural networks (ANN; \citealt{Whitten2019, Ksoll2020, Whitten2021}), have been suggested to be effective ways to derive precise atmospheric parameters from photometric colors. In particular, ANNs that build an explicit function to map photometric colors to stellar parameters have become a popular tool for estimating atmospheric parameters and elemental abundances.
With the J-PLUS photometry, \citet{Whitten2019} proposed SPHINX, a network of ANNs, to derive $T_{\rm eff}$ and [Fe/H] over the range of 4500\,K $< T_{\rm eff}<6200$\,K, and obtain [Fe/H] down to about $-$3.0 in J-PLUS DR1\footnote{\url{http://www.j-plus.es/datareleases/data_release_dr1}}, with a typical scatter of $\delta$ [Fe/H] $=$ 0.25 dex.
Later, \citet{Whitten2021} extended their ANN approach to estimate [C/Fe] to a precision of $\sim 0.35$ dex for SDSS Stripe 82 stars contained in S-PLUS DR2 \citep{Almeida-Fernandes2021} . 

The J-PLUS narrow-band ($\sim 100$\,\AA) filters are centered on key stellar absorption
features; $J0378$ for the CN band, $J0395$ for Ca\,II H+K, $J0410$ for H$\delta$, $J0430$ for the CH $G$-band, $J0515$ for the Mg $b$ triplet, $J0660$ for H$\alpha$, and $J0861$ for the Ca triplet. 
Such specially designed filters make it possible to not only determine the basic stellar atmospheric 
parameters ($T_{\rm eff}$, $\log g$, and [Fe/H]), but also to constrain  [$\alpha$/Fe] and elemental abundances such as [C/Fe], [N/Fe], [Mg/Fe] and [Ca/Fe]. 
Motivated by the above possibilities and their high scientific impact,  we have developed a cost-sensitive set of neural networks, {\it CSNet}, to derive precise and robust stellar labels ($T_{\rm eff}$, $\log g$, [Fe/H], [C/Fe], [N/Fe], [Mg/Fe], [Ca/Fe],and [$\alpha$/Fe]) for stars in J-PLUS DR1, adopting J-PLUS stars in common with LAMOST and {\it Gaia} as training sets. Our models improve the prediction accuracy on the whole by increasing the error penalty for the relatively rare samples of extreme stars.

The paper is organized as follows: Section\,\ref{data} describes the data used in this work. Section \,\ref{method} introduces the framework of the proposed method in detail. Section \,\ref{Result} reports the results for the training and the testing samples, as well as comparisons with various validation samples. The resulting catalog of stellar parameters and 
chemical abundances for stars in JPLUS-DR1 are also presented in Section \,\ref{Result}.
Section \,\ref{discussion} discusses some challenges associated with this work, followed by a summary in Section \,\ref{summary}.

\section{Data} \label{data}
The dataset used for training and testing {\it CSNet} is constructed by stars in common between J-PLUS DR1 (\citealt{Cenarro2019}), {\it Gaia} DR2 (\citealt{Brown2018}), and LAMOST DR5 (http://dr5.lamost.org, \citealt{Luo2015,Xiang2019}), where the first two data sets provide input stellar colors and the last one provides stellar labels ($T_{\rm eff}$, $\log g$, [Fe/H], [C/Fe], [N/Fe], [Mg/Fe], [Ca/Fe] and [$\alpha$/Fe]). 

\subsection{Stellar colors} \label{J-PLUS DR1}

J-PLUS\footnote{\url{www.j-plus.es}} is being conducted from the Observatorio Astrof\'{\i}sico de Javalambre (OAJ, Teruel, Spain; \citealt{Cenarro2014}) using the 83\,cm Javalambre Auxiliary Survey Telescope (JAST80) and T80Cam, a panoramic camera of 9.2k $\times$ 9.2k pixels that provides a $2\deg^2$ field of view (FoV) with a pixel scale of 0.55 arsec pix$^{-1}$ \citep{Marin2015}. Its unique combination of 5 broad-band ($ugriz$), and 7 medium- and narrow-band filters ($J0378, J0395, J0410, J0430, J0515, J0660, J0861$), optimally designed to extract the rest-frame spectral features (including the Balmer jump region, which covers the molecular CN feature, Ca\,II H+K, H$\delta$, the CH $G$-band, the Mg $b$ triplet, H$\alpha$, and the Ca triplet) plays a key role in characterizing stellar types and elemental abundances in this work. The J-PLUS observational strategy, image reduction, and main 
scientific goals are presented in \citet{Cenarro2019}. 

The first J-PLUS Data Release (DR1) covers 1022 $\text{deg}^2$ with 511 pointings in its footprint observed from November 2015 to January 2018 (\citealt{Cenarro2019}). 
The 5$\sigma$ limiting magnitudes (3$''$ aperture) in the filters reach a limit of about 21.
Using a stellar locus method, \citet{Sanjuan2019} have reached a calibration accuracy of 1-2 per cent, larger for bluer filters.
Using the stellar color regression method (SCR; \citealt{Yuan2015a}, \citealt{Huang2021b}, 
\citealt{Niu2021a}, \citealt{Niu2021b}, Huang \& Yuan 2021), by combining the LAMOST DR5 spectroscopy and {\it Gaia} DR2 photometry, 
we have re-calibrated the J-PLUS DR1 data and achieved an accuracy of about 0.5 per cent or better for all the J-PLUS filters (Yuan et al. in preparation).
Strongly correlated calibration errors are found in the previous calibration, due to the strong metallicity-dependence of stellar loci for blue filters (e.g., \citealt{Yuan2015b}, \citealt{Sanjuan2021})
and errors in the 3D dust map and reddening coefficients used (Yuan et al. in preparation).
In order to make the best use of the full power of J-PLUS filters, the re-calibrated J-PLUS DR1 data are used in this work. 
The same data have also been used in star/galaxy/quasar classifications (Wang et al. 2021a), as well as for estimates of basic stellar parameters (Wang et al. 2021b), with a Support Vector Machine technique.

{\it Gaia} DR2 (\citealt{Brown2018}) delivers five-parameter astrometric solutions 
as well as integrated photometry in three very broad bands: $G$, $B$P (330 -- 680 nm), and $RP$ (630 -- 1 050 nm), for 1.4 billion sources with $G$ $<$ 21. 
It provides not only the best astrometric data ever obtained,  but also the most precise photometric data. 
The typical uncertainties in {\it Gaia} DR2 measurements at $G$ = 17 are $\sim$ 2 mmag in the $G$-band photometry, and $\sim$ 10 mmag in $BP$ and $RP$ magnitudes 
(\citealt{Brown2018}).

With the additional $BP$ and $RP$ magnitudes from {\it Gaia} DR2, a total of 13 stellar colors are computed and used in this work, as listed in Table\,\ref{table:table1}. These combinations of colors are believed to contain all the pertinent stellar parameter and elemental-abundance information hidden in the J-PLUS photometry. Reddening corrections have been applied to these colors with empirical reddening coefficients determined with the star-pair technique (\citealt{Yuan2013}; Yuan et al. in preparation) and $E(B-V)$ reddening 
values from the SFD98 map (\citealt{Schlegel1998}). The coefficients are also listed in Table\,\ref{table:table1}. 

\begin{table}
\centering
\begin{threeparttable}[b]
\caption{Empirical reddening coefficients for stellar colors.}
\label{table:table1}
\begin{tabular}{llc}
\hline
 No. &  Colors$^1$ &  Empirical reddening coefficients \\
\hline 
  1 &  $BP-RP$ &  $\,\,\,\,1.360$ \\
  2 &  $BP-u$ &  $-1.379$ \\
  3 &  $BP-g$ &  $-0.339$ \\
  4 &  $RP-r$ &  $-0.678$ \\
  5 &  $RP-i$ &  $-0.033$ \\
  6 &  $RP-z$ &  $\,\,\,\,0.439$ \\
  7 &  $BP-J0378$ &  $-1.203$ \\
  8 &  $BP-J0395$ &  $-1.158$ \\
  9 &  $BP-J0410$ &  $-0.953$ \\
  10 &  $BP-J0430$ &  $-0.809$ \\
  11	&  $BP-J0515$ &  $-0.069$ \\	
	12	&  $RP-J0660$ &  $-0.454$ \\
	13 &  $RP-J0861$ &  $\,\,\,\,0.339$ \\ 
\hline
\end{tabular}
\begin{tablenotes}
  \item[1]{$BP$ and $RP$ photometry are from {\it Gaia} DR2;
  $ugriz$, $J0378$, $J0395$, $J0410$, $J0430$, $J0515$, $J0660$, and $J0861$ photometry are from J-PLUS DR1.}
\end{tablenotes}
\end{threeparttable}
\end{table}

The input colors for {\it CSNet} are constructed by cross-matching J-PLUS DR1 with {\it Gaia} DR2, adopting a matching radius of 1.0 arcsec. 
Considering that the 6 arcsec aperture measurements of J-PLUS DR1 are used, we require that stars satisfy {photo\_bp\_rp\_excess\_factor}  $\le$ 1.25 + $0.06(BP-RP)^2$, slightly stricter than 
that suggested by \citet{Evans2018}, to avoid possible contaminations from nearby sources. To ensure the quality of the photometry, we require that stars satisfy FLAGS $=$ 0. For the training and the testing samples, we further require that photometric uncertainties of the 12 J-PLUS filters are lower than 0.01 mag for the 
{\it griz} filters, 0.02 mag for the $J0410, J0430$ and $J0515$ filters, and 0.03 mag for the $J0378, J0395$, and $J0410$ filters, respectively. With the above contraints, the G magnitude ranges over 11.6 -- 16.7 for the training and the testing samples.

\subsection{LAMOST stellar labels} \label{LAMOST DR5}
The Large sky Area Multi-Object fiber Spectroscopy Telescope (LAMOST; \citealt{Cui2012}) collects low-resolution (R$\sim$1800) and medium-resolution (R$\sim$7500) spectra in a FoV of 20 $\text{deg}^2$. 
In its fifth Data Release, LAMOST DR5, this survey has delivered more than 8 million stellar spectra with spectral resolution $R = 1800$ and limiting magnitude of $r \sim$\,17.8  (\citealt{Deng2012,Liu2014}). Stellar effective temperatures, $T_{\rm eff}$, surface gravities, $\log g$, and metallicities, [Fe/H], are derived by the LAMOST Stellar Parameter Pipeline (LASP; \citealt{Wu2011}). 

For the elemental abundances [C/Fe], [N/Fe], [Mg/Fe], [Ca/Fe], and [$\alpha$/Fe], the data-driven {\it Payne} ({\it The DD-Payne}) results of \citet{Xiang2019} are used. 
{\it The DD-Payne} inherits essential ingredients from both {\it The Payne} (\citealt{Ting2019}) and {\it The Cannon} (\citealt{Ness2015}), and 
incorporates constraints from theoretical model spectra to ensure physically meaningful abundance estimates. 
Stars in common between LAMOST DR5 and either GALAH DR2 (\citealt{Buder2018})  or APOGEE DR14 (\citealt{Holtzman2018}) were used as the training data set to provide abundances\footnote{\url{http://dr5.lamost.org/doc/vac}} 
for 16 elements (C, N, O, Na, Mg, Al, Si, Ca, Ti, Cr, Mn, Fe, Co, Ni, Cu, and Ba).
For stars with spectral signal to noise ratios $\rm S/N_g$ higher than 50, the typical internal uncertainties of the estimated abundances are 
about 0.05 dex for [Fe/H], [Mg/Fe], and [Ca/Fe], 0.1dex for [C/Fe] and [N/Fe].
The [$\alpha$/Fe] of this catalog was defined as a weighted mean of [Mg/Fe], [Ca/Fe], [Ti/Fe], and [Si/Fe]. 
Note that for the elemental abundances used in this work, [Mg/Fe] was trained using GALAH DR2, while the others used APOGEE DR14.   
         
\subsection{Experimental data construction} \label{data construction}

\begin{figure*}
  \centering
  \includegraphics[width=\textwidth]{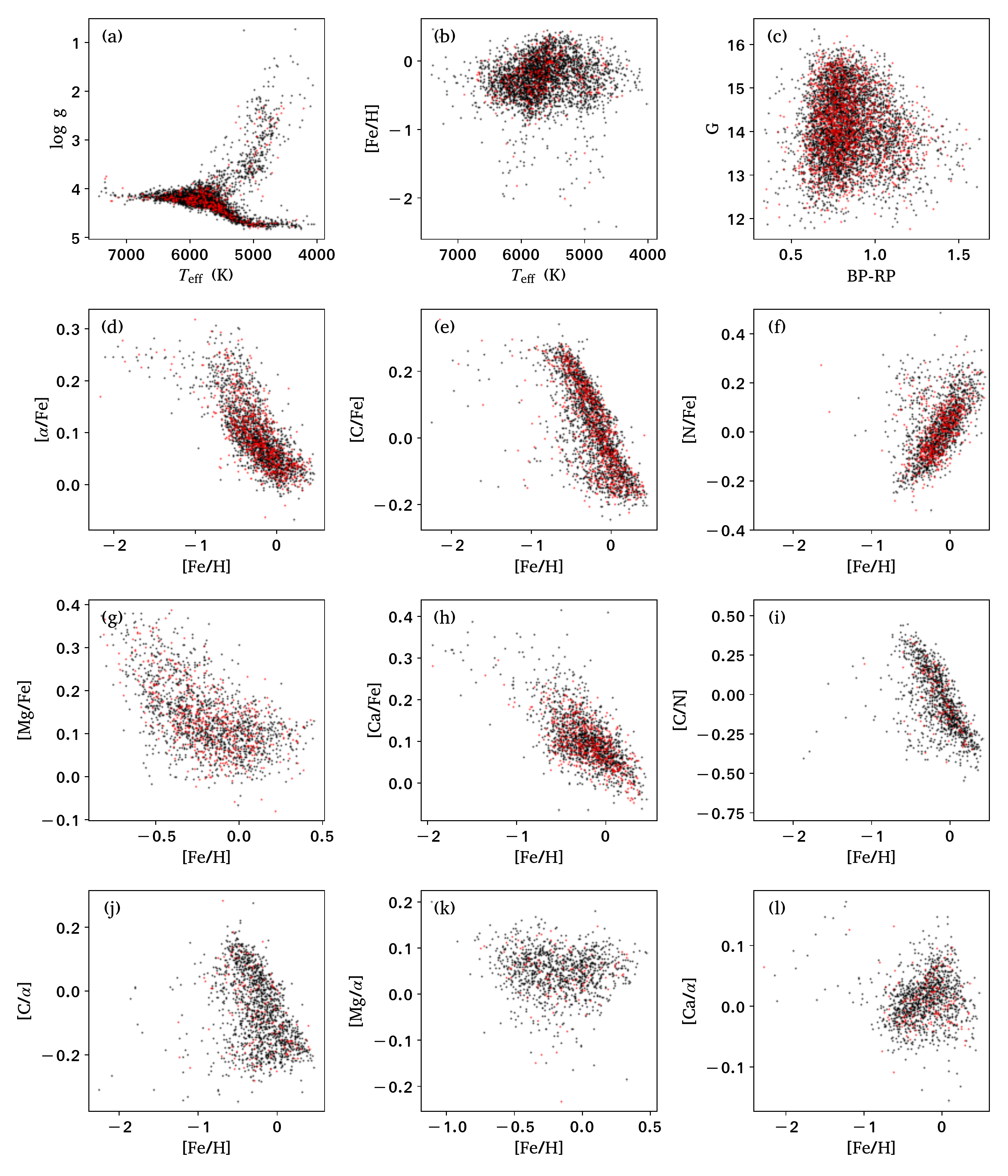}
  \caption{Distributions of the {\it CSNet} training and the testing samples in the planes of $T_{\rm eff}$--$\log g$, $T_{\rm eff}$--[Fe/H], [$BP$--$RP$]--$G$, [$\rm \alpha$/Fe]--[Fe/H], [C/Fe]--[Fe/H], [N/Fe]--[Fe/H], [Mg/Fe]--[Fe/H], [Ca/Fe]--[Fe/H], [C/N]--[Fe/H], [C/$\alpha$]--[Fe/H], [Mg/$\alpha$]--[Fe/H], and [Ca/$\alpha$]--[Fe/H], from top to bottom and left to right. The black and red dots represent the training set and testing set stars, respectively. To avoid crowding, only 10 per cent of selected stars are plotted.}
  \label{fig:Fig3}
\end{figure*}

The experimental data set for training and testing {\it CSNet} consists of 67,709 stars within certain constraints for the {photo\_bp\_rp\_excess\_factor} and the photometric uncertainties (see Section \ref{J-PLUS DR1} for more details). In particular, eight {\it CSNet}s with the same structure and hyper-parameters are trained: three for basic stellar atmospheric parameters and five for elemental abundances. Considering the density distribution of the experimental data set in the label space, we selected the experimental stars with extra criteria shown in Table\,\ref{table:table2}.
These stars were divided into the training set and the testing set in the ratio of 3:1. Fig.\,\ref{fig:Fig3} shows the distributions of the training and the testing sets in the planes of $T_{\rm eff}$--$\log g$, $T_{\rm eff}$--[Fe/H], [$BP$--$RP$]--$G$, [$\rm \alpha$/Fe]--[Fe/H], [C/Fe]--[Fe/H], [N/Fe]--[Fe/H], [Mg/Fe]--[Fe/H], [Ca/Fe]--[Fe/H], [C/N]--[Fe/H], [C/$\alpha$]--[Fe/H], [Mg/$\alpha$]--[Fe/H], and [Ca/$\alpha$]--[Fe/H]. We can see that both the training set and the testing set have wide coverages in temperature and surface gravity: 4000\,K $<T_{\rm eff} < 7500$\,K and $0.0< \log g < 5.0$. Panel (g) of Fig.\,\ref{fig:Fig3} shows that stars with reference [Mg/Fe] values have metallicities in the range $-1.0 < $ [Fe/H] $< +0.5$. Since the training and the testing sets were randomly selected from the original (trimmed) dataset, both have similar distributions. Note that panel (a) of Fig.\,\ref{fig:Fig3} shows no stars with $\log g < 2.0$ in the testing set. This is simply due to the 10 per cent random selection procedure applied to the data.

\begin{table*}
  \centering
  \caption{Adopted contraints on the datasets for training and testing {\it CSNet}.}
  \begin{threeparttable}
  \label{table:table2}
  \begin{tabular}{lllcl}
  \hline
  \multirow{2}{*}{Parameters} 	& \multirow{2}{*}{Number}  & \multicolumn{3}{c}{Constraint}   \\
  \cline{3-5} 							    &                 & Effective parameter range & Effective $BP$--$RP$ range & $\rm S/N_g$  \\
  \hline
  $\textit{T}_{\text{eff}}$\tnote{1} 		&56221  & 4000\,K $<\textit{T}_{\text{eff}}<7500 \rm \,K$  &[0.319,1.660]   & $>20$    	\\
  $\log g$\tnote{1}        						&56360  & $0.00<\log g<5.00$  &[0.062,1.786]  & $>20$    	\\
  $\text{[Fe/H]}$\tnote{1}          		&56322  & $-2.50<\text{[Fe/H]}<+0.50$ &[0.062,1.786] & $>20$     	\\
  $\text{[}$$\alpha$$\text{/Fe]}$\tnote{2}    &25108  &$-0.10<\text{[}$$\alpha$$\text{/Fe]}<+0.40$ &[0.062,1.716] & $>50$      	\\
  $\text{[C/Fe]}$\tnote{2} 						&24824  &$-0.30<\text{[C/Fe]}<+0.40$ &[0.289,1.716]  &$>50$    	\\
  $\text{[N/Fe]}$\tnote{2}						&18714  &$-0.50<\text{[N/Fe]}<+0.50$ &[0.527,1.716] &$>50$     		\\
  $\text{[Mg/Fe]}$\tnote{2}						&18714  &$-0.10<\text{[Mg/Fe]}<+0.40$ &[0.364,1.586]  &$>50$    	\\
  $\text{[Ca/Fe]}$\tnote{2}						&18714  &$-0.15<\text{[Ca/Fe]}<+0.50$ &[0.315,1.626]  &$>50$    		\\
  \hline
  \end{tabular}
  \begin{tablenotes}
  \item[1] Stellar parameters of the reference catalog from LAMOST DR5.
  \item[2] Elemental abundances of the reference catalog from the {\it DD--Payne} results.
  \end{tablenotes}
  \end{threeparttable}
\end{table*}

\section{Method: The cost-sensitive ANN} \label{method}

Considering the nature of the photometric data from J-PLUS DR1, we have developed {\it CSNet}, a combination of a traditional ANN architecture and a novel two-dimensional cost-sensitive learning algorithm to estimate stellar parameters and chemical abundances. Training the neural network is performed by the cost-sensitive learning algorithm in order to achieve better measurement precision.

\subsection{Data normalization} \label{normalize}

Input variables with the same scale are the basis for training the robust {\it CSNet}. Thus, stellar colors are standardized before entering the network by the following z-score normalization:
\begin{equation}
  {x}'=\frac{x-\mu }{\sigma},
\end{equation}
where $x$ and ${x}'$ are the original and standardized input vectors, respectively, with the 13 stellar colors respectively, while $\mu$ and $\sigma$ are the mean and standard deviation of all the original input vectors, respectively.

\subsection{The ANN} \label{ANN}

\begin{figure*}
  \centering
  \includegraphics[width=\textwidth]{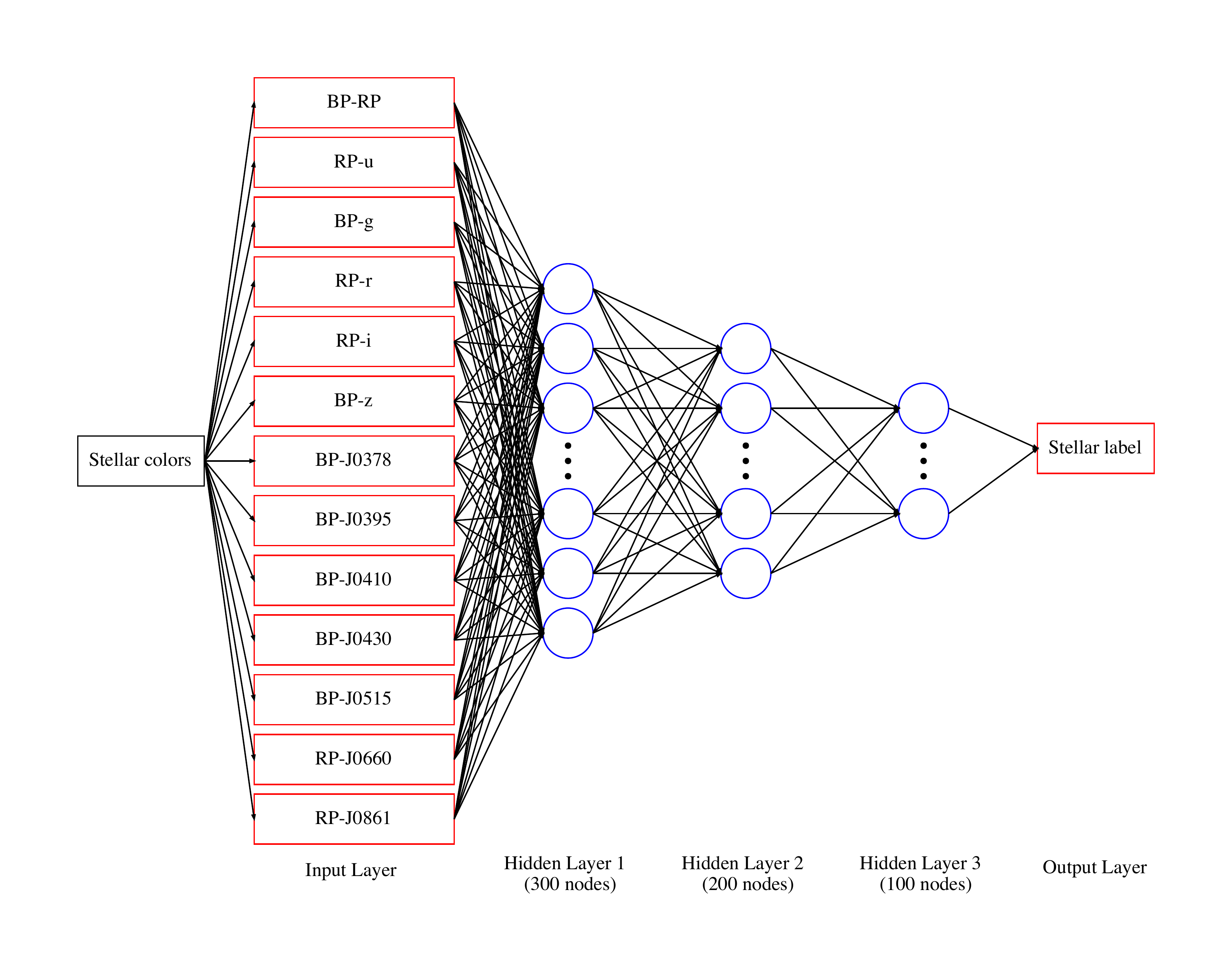}
  \caption{The ANN structure of {\it CSNet}. The structure consists of a scalable five-layered feed-forward network. The input layer imports the 13 stellar colors transformed by J-PLUS DR1 and {\it Gaia} DR2 after extinction correction as the basic features. The following three hidden layers with activation functions extract the deep non-linear features of the stellar colors. The output layer provides a stellar label with a weighted sum of the learned features in the last hidden layer.}
  \label{fig:Fig2}
\end{figure*}

To balance the under-fitting and over-fitting during the training process,
after several experiments, the appropriate architecture for the multilayered feed-forward network in this work is 13-300-200-100-1 for estimating each stellar label, consisting of three hidden layers to extract deep features from the photometry with the 13 stellar colors. The model structure is illustrated in Fig.\,\ref{fig:Fig2}.

Each node calculates a weighted sum of all its input values and produces an output value with a non-linear activation function. This process can be described as:
\begin{gather}
	H^l=g(w^{(l)}X^l+b^{(l)}),\\
  X^{(l+1)}=H^l,
\end{gather}
where $X^l$ and $H^l$ represent the input vector and output vector of the $l$-th layer respectively, $w$ is the weight vector, $b$ represents the bias terms and $g(\bullet)$ represents the LeakyReLU activation function with negative slope coefficient $\alpha = 0.01$ in the hidden layer to solve “Dead Neuron” problems.

Then, adjustable parameters (including $w$ and $b$) of the ANN can be calculated through minimizing the following cost:
\begin{equation}
  C(w,b)=\left \| \hat{Y}-\sum_{l=1}^{L}g_{w^{(l)},b^{(l)}}(X^l)\right \|_2^2,
\end{equation}
where $ \hat{Y}$ denotes the corresponding known stellar labels.

However, the over-fitting problem is more likely to occur when a large number of neurons are employed to extract deep features from the input data, which means that the accuracy of the prediction in the testing set decreases as the performance of the model in the training set improves. $L_2$ regularization, Dropout layers and Batch Normalization layers are all effective to mitigate this problem. In this work, a $L_2$ regularized term is added to the cost function above as follows:
\begin{equation} 
  \label{cost function}
  C(w,b)=\left \| \hat{Y}-\sum_{l=1}^{L}g_{w^{(l)},b^{(l)}}(X^l)\right \|_2^2+\lambda \left \| w\right \|_2^2,
\end{equation}
where $ \lambda \in (0,1)$ is the trade-off coefficient between the residual error and the regularized term.

The back-propagation learning algorithm with adaptive moment estimation (ADAM) is utilized to minimize the cost function Eq.\,(\ref{cost function}) effectively. Let $\theta =(w,b)$ be parameters of the model. Then, the change of parameter $\theta$ becomes:
\begin{equation} 
  \Delta \theta (t)=-\frac{\eta }{2} \nabla_{\theta} C(\theta (t-1))+\alpha \Delta \theta (t-1),
\end{equation}
where $\eta$ is the step size, $t$ is the number of current iterations, and $ \alpha \in [0,1]$ is the decay rate of the previous weight change.

The process of obtaining the resulting parameters $\theta$ can be described as follows (see \citealt{Adam}):

\textbf{Initialize:} $ \eta _t=\frac{\eta }{\sqrt{t}} $ as the step size, $\beta _1, \beta _2\in (0,1)$ as the decay rates for the moment estimate, $ \epsilon >0$, $ C(\theta (t))$ as a convex differentiable cost function, $ \theta _0$ as an initial parameter vector, $m_0$ and $v_0$ as an initial 1st and 2nd moment moment vector, respectively, and $t=0$ as the initial time step.

\textbf{while:} $ \theta (t) $ not converged \textbf{do:} 
  
\,\,\,\,\,\,$t=t+1$

\,\,\,\,\,\,$g_t=\nabla_\theta C(\theta (t-1))$

\,\,\,\,\,\,$m_t=\beta _1 m_{t-1}+(1-\beta _1)g_t$

\,\,\,\,\,\,$v_t=\beta _2 v_{t-1}+(1-\beta _1)g_t^2$

\,\,\,\,\,\,$\hat{m_t}=m_t/(1-\beta _1^t)$

\,\,\,\,\,\,$\hat{v_t}=v_t/(1-\beta _2^t)$

\,\,\,\,\,\,$\theta (t)=\theta (t-1)-\eta _t\hat{m_t}/(\sqrt{\hat{v_t}}+\epsilon )$

\textbf{end}

\textbf{return} $\theta (t)$

Next, a mapping from the input stellar colors to the output stellar labels is established with definite parameters $\theta =(w,b)$.

\subsection{Modifications of the ANN} \label{CS-learning}

When the training set exhibits an unbalanced distribution in the population, the most frequent cases dominate the predicted values. Namely, the predictions will have a systematic trend towards the coverage of the greatest number of target values. \citet{Oommen2011} showed that techniques reducing the sampling bias in the target value space could improve the prediction precision for regression tasks. 

In this paper, we modify the cost function in the ANN which takes the rare target values into account. Given an input dataset $X$ with continuous numeric label pairs $(Y_1,Y_2)$, a frequency distribution histogram is generated with $M*N$ bins in $Y_1$--$Y_2$ space, where $Y_1$ represents one of the given stellar labels and $Y_2$ represents the [Fe/H] value. Next, different costs are computed for samples in different bins of the histogram with the following rule:
\begin{equation} \label{frequency distribution histogram}
  c(x_i)=\left(\frac{f_n(y_1^i,y_2^i)}{max(f_n(Y_1.Y_2))}\right)^{-\gamma},
\end{equation}
where $x_i\in X$ is a sample, $f_n(\bullet)$ is a function to calculate the frequency of samples in a bin which includes $(y_1^i,y_2^i)$ in the histogram, and $\gamma>0$ controls the difference degree of the cost among different bins.

Then, the cost function of ANN in Eq.\,(\ref{cost function}) is changed to:
\begin{equation} 
  \label{final cost function}
  C(w,b)=c(X)\left \| \hat{Y}-\sum_{l=1}^{L}g_{w^{(l)},b^{(l)}}(X^l)\right \|_2^2+\lambda \left \| w\right \|_2^2,
\end{equation}
This favors labels of the minority cases with higher expected error costs.

\section{Results} \label{Result}

To demonstrate the accuracy and reliability of the results from  {\it CSNet}, we perform extensive experiments examining the estimated stellar labels from different aspects. After introducing the experimental setting, we train and test {\it CSNet} with the constructed data set (see Section\,\ref{data construction}), and compare the measurements on the stars in common with other precision survey catalogs. Then we apply this model to 4,387,568 stars (MAGABDUALOBJ\_CLASS\_STAR $\geq$ 0.6) selected by cross-matching J-PLUS DR1 with {\it Gaia} DR2.

\subsection{Experimental setting} \label{Experimental setting}

The experiments are performed with Keras 2.1.4, Tensorflow 1.5.0, CUDA 9, and cuDNN 7. ADAM is used as the optimizer because of its good robustness to the initial learning rate. Before training {\it CSNet}, there are several hyper-parameters that have been set manually: $s$ is the batch size of the training samples, $epoch$ defines the training iterations, $\eta$ is the learning rate, $M*N$ is the number of bins in the training set frequency distribution histogram, $\gamma $ is the exponent of the weight assignment function Eq.\,(\ref{frequency distribution histogram}), $\lambda $ is the penalty coefficient of the cost function Eq.\,(\ref{final cost function}), and the parameters ($\beta _1, \beta _2, \epsilon $) for ADAM. Large $s$ occupying the GPU memory speeds the convergence, small $\eta$ consuming more time helps find the optimal value, proper $\gamma$ ensures similar distribution between the training and testing set, $\lambda$ is positively associated with the typical value of the loss function, and other hyper-parameters ($\beta _1, \beta _2, \epsilon $) in ADAM can use its default values. From experimentation, we found that reasonable default settings for the stellar label determination are $s=2000$, $epoch=5000$, $\eta=10^{-4}$, $\gamma=0.5$, $\lambda=5*10^{-5}$, $\beta_1=0.9$, $\beta_2=0.999$, and $\epsilon=10^{-8}$. According to the range of the expected stellar labels, recommended bins (M*N) are ($20*1$), ($20*1$), ($20*1$), ($20*60$), ($28*60$), ($35*50$), ($20*40$) and ($25*60$) for $T_{\rm eff}$, $\log g$, [Fe/H], [$\alpha$/Fe], [C/Fe], [N/Fe], [Mg/Fe], and [Ca/Fe], respectively. 

\subsection{Performance} \label{Performance}

We restrict the application of {\it CSNet} only to stars within the same [$BP$--$RP$] coverage as the training set. Furthermore, only stellar label estimates located in the same range with the training set (see Table\,\ref{table:table2}) are considered to be reliable, as we are cautious to extrapolate, a well-known limitation of ANN-based approaches. We evaluated the performance of these labels on the training, testing, and validation samples.

\subsubsection{Parameter determination on the training and the testing sets}\label{performance of experimental data}

To assess the accuracy of stellar labels derived from our model, we first compare the predictions from the model and their corresponding LAMOST labels in both the training and the testing sets. 

\begin{figure*}
  \centering
  \includegraphics[width=\textwidth]{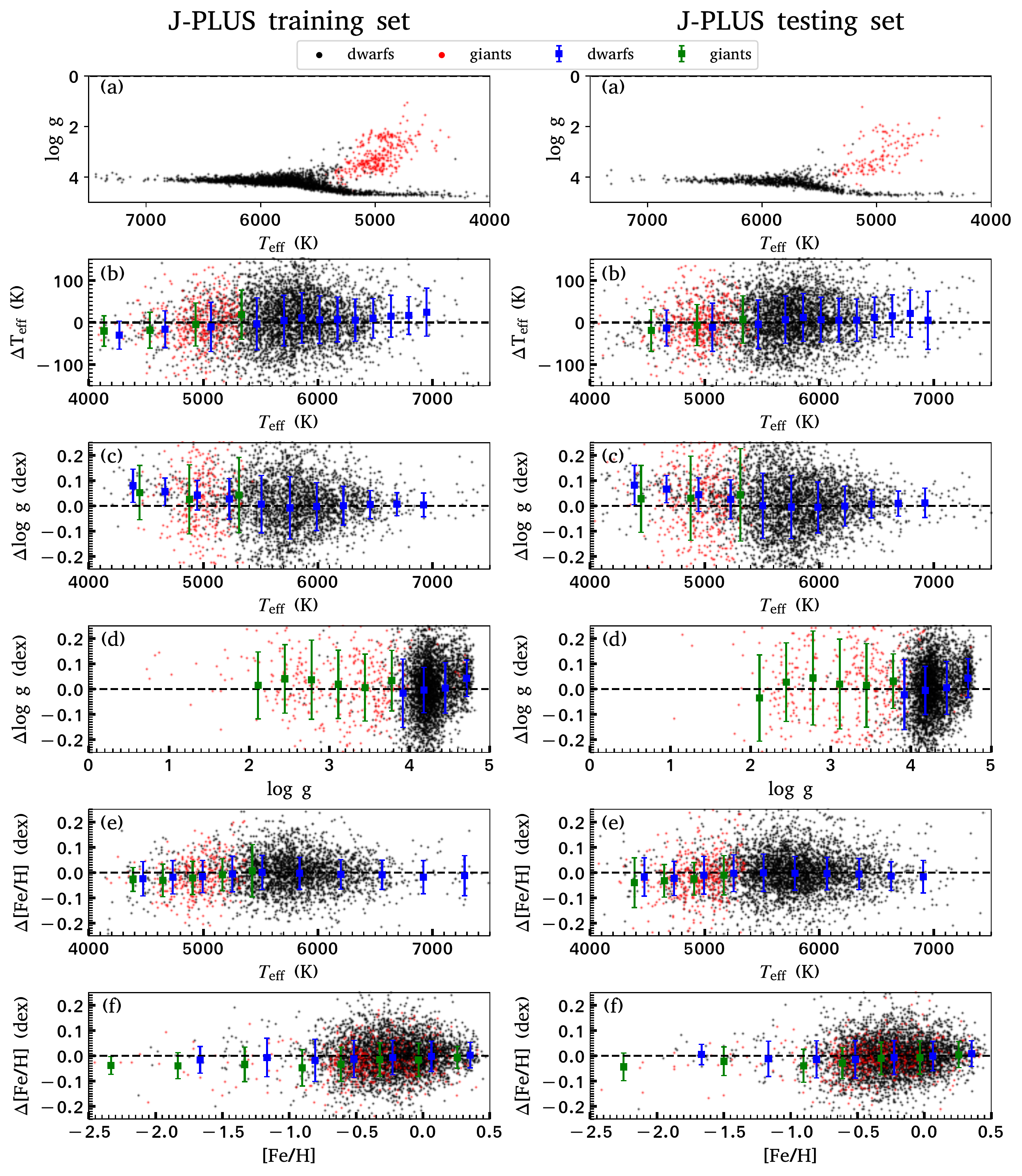}
  \caption{Training set (left) and testing set (right) for  $T_{\rm eff}$, $\log g$, and [Fe/H] for the {\it CSNet} results. The black and red dots represent dwarfs and giants, respectively. Error bars, colored by blue and green for dwarfs and giants respectively, denote the mean value ``bias” and 1 $\sigma$ uncertainty of the residuals estimated using Gaussian fits.}
  \label{fig:lamost_parameter}
\end{figure*}

For the stellar parameters ($T_{\rm eff}$, $\log g$, and [Fe/H]), Fig.\,\ref{fig:lamost_parameter} plots distributions of the {\it CSNet} results in the plane of $T_{\rm eff}$--$\log g$ (panel (a)) and residuals for dwarfs and giants as a function of effective temperature (panels (b), (c), and (e)), surface gravity (panel (d)) and metallicity (panel (f)). Giant stars ($BP-RP>0.95$ and $M_G<3.9$) and dwarf stars ($BP-RP\leq 0.95$ or $M_G\geq 3.9$) are distinguished in the color-magnitude diagram. The top two panels show that giants and dwarfs are also clearly distinguished in the plane of $T_{\rm eff}$--$\log g$. Errors of the predicted stellar labels are evaluated by the mean deviation (``bias”) and $1\sigma$ uncertainties, which are estimated using Gaussian fits. There is no significant bias in the training and the testing sets between the {\it CSNet} results and the LAMOST values, even in regions with small numbers of training stars, demonstrating that the trained model is robust to the small sample field and avoids over-fitting. The uncertainties of the residuals are $\delta\, T_{\rm eff}\sim$ 55\,K, $\delta \log g\sim 0.15$\,dex, and $\delta$ [Fe/H] $\sim$ 0.07\,dex, respectively. 

\begin{figure*}
  \centering
  \includegraphics[width=\textwidth]{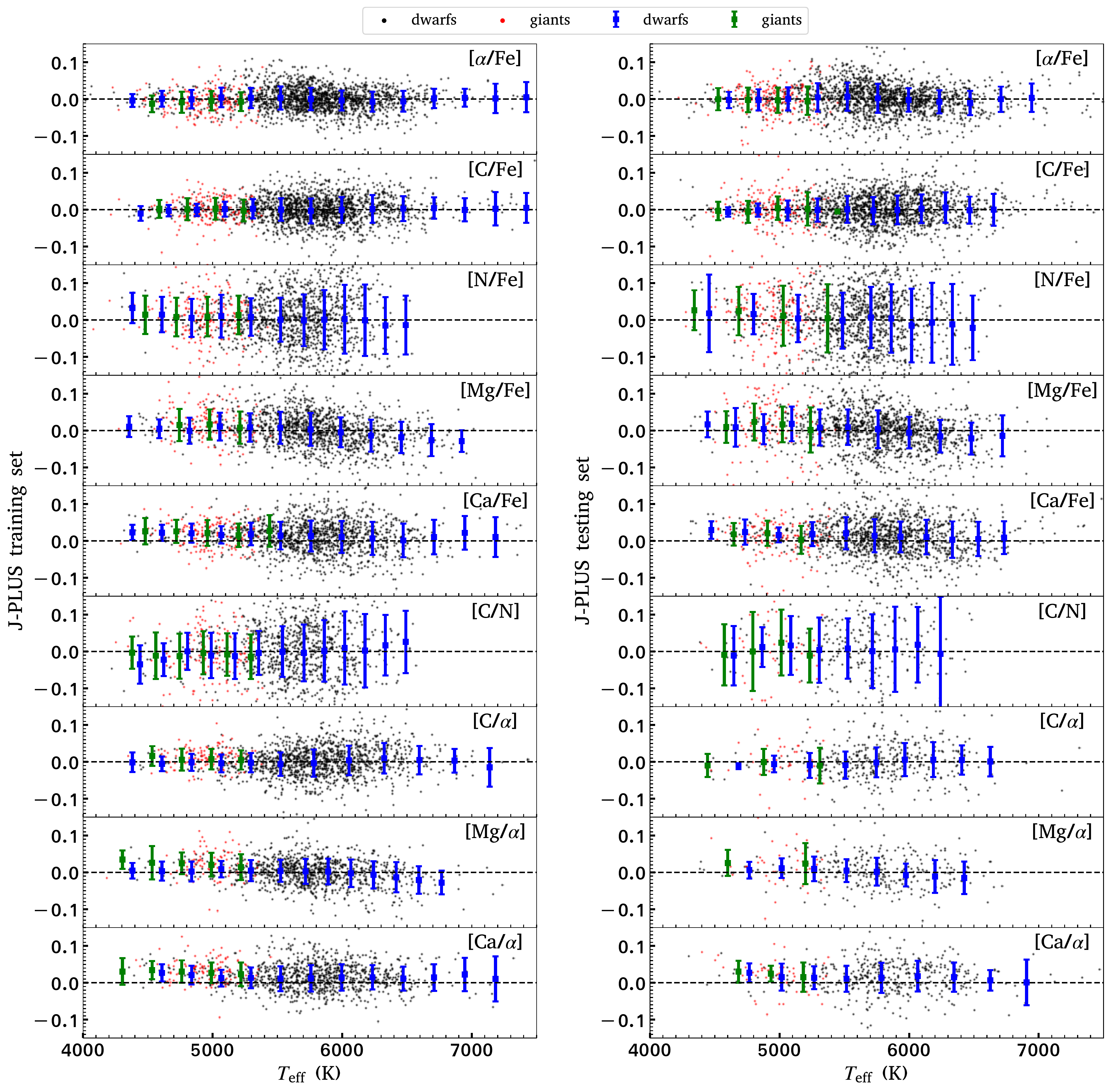}
  \caption{Similar to Fig.\,\ref{fig:lamost_parameter}, but here we show residuals for the {\it CSNet} elemental abundances as a function of effective temperature.}
  \label{fig:lamost_abundance}
\end{figure*}

Similarly, for the elemental abundances, Fig.\,\ref{fig:lamost_abundance} indicates that the results are commensurate with the reference values in both the training and the testing sets. The uncertainties of the residuals are $\delta\,$[$\alpha$/Fe] = 0.03\,dex, $\delta$\,[C/Fe] = 0.04\,dex, $\delta$\,[N/Fe] = 0.08\,dex, $\delta$\,[Mg/Fe] = 0.05\,dex, and $\delta$\,[Ca/Fe] = 0.05\,dex, respectively. The low levels of bias and uncertainty between the {\it CSNet} predictions and LAMOST {\it DD-Payne} values suggest that the trained model performs well on abundance determinations over the validity range.

\subsubsection{Comparisons with other surveys} \label{Comparison}

To examine the reliability of the {\it CSNet} results, we select stars in common between the J-PLUS DR1 and other reference catalogs including precise stellar parameters and elemental abundances derived from their available spectra.
\begin{enumerate}
  \item{APOGEE--{\it Payne}}: The Apache Point Observatory for Galactic Evolution Experiment (APOGEE; \citealt {Majewski2017}) is a high-resolution ($R\sim22,500$), high signal-to-noise ratio ($\rm S/N>100$), near-infrared (1.51--1.70\,$\mu$m) spectroscopic survey. \citet{Ting2019} provide accurate stellar parameters and abundances derived from the APOGEE DR14 spectra with the neural-network based method {\it The Payne}. We cross-match the J-PLUS DR1 stars with the APOGEE--{\it Payne} catalog, finding 7,703 stars in common.

  \item{APOGEE--{\it ASPCAP}}: APOGEE Stellar Parameters and Chemical Abundances Pipeline (ASPCAP; \citealt {Perez2016}) produces a catalog of stellar labels for the Data Release 15 (DR15) of APOGEE with all quantities for each combined spectrum. We cross-match the J-PLUS DR1 with the APOGEE--{\it ASPCAP} catalog for stars with S/N higher than 100, and obtain 10,688 stars in common.
  
  \item{GALAH--{\it Cannon}}: The Galactic Archaeology with HERMES (GALAH; \citealt{De_Silva2015}) is a 
  high-resolution ($R\sim28,000$) spectroscopic survey using the Anglo-Australian Telescope over a 2 degree field of view. The Data Release 2 (DR2) of GALAH contains the catalog of stellar parameters and abundances determined by the data-driven algorithm {\it The Cannon} (\citealt {Ness2015}). We cross-match the J-PLUS DR1 with the GALAH--{\it Cannon} catalog and obtain 252 stars in common.  
   
\end{enumerate}

\begin{figure*}
  \centering
  \includegraphics[width=\textwidth]{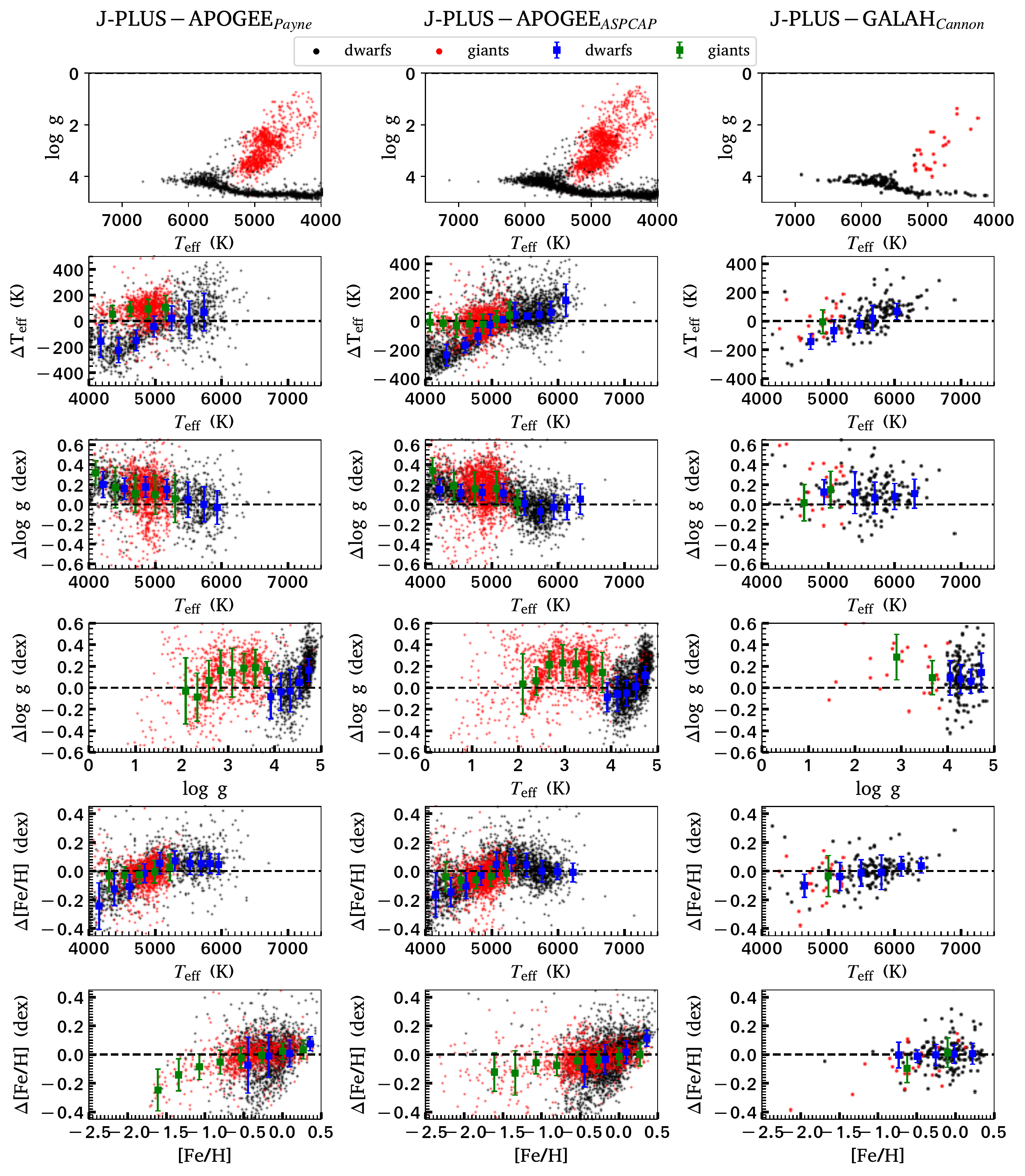}
  \caption{Similar to Fig.\,\ref{fig:lamost_parameter}, but for stars in common between the {\it CSNet} results and the reference catalogs.}
  \label{fig:cross_parameter}
\end{figure*}

Fig.\,\ref{fig:cross_parameter} shows the comparisons of $T_{\rm eff}$, $\log g$, and [Fe/H] between our results and the above three reference catalogs. Overall, our results are in good agreement with the values from these  catalogs. We do not expect a perfect match, considering that the reference LAMOST DR5 catalog used for training has differences with respect to these three catalogs. The APOGEE--{\it Payne} $T_{\rm eff}$ values are consistent with ours for both giant and dwarf stars, except for dwarfs with $T_{\rm eff}$ lower than 4800\,K, where there is an obvious systematic trend --- the APOGEE--{\it Payne} values are higher than ours, and the bias reaches about 100--220\,K. Another noticeable difference is for the giant stars with $T_{\rm eff}$ lower than 4300\,K, as the APOGEE--{\it Payne} $\log g$ values are lower than ours, and the bias reaches 0.2--0.3\,dex. The difference between [Fe/H] from our result and that from the APOGEE--{\it Payne} shows weak systemic trends for dwarf stars with $T_{\rm eff}$ lower than 4800\,K, and the bias reaches 0.10--0.22\,dex. Similar to the comparisons between the APOGEE--{\it Payne} values and ours, APOGEE--{\it ASPCAP} and GALAH--{\it Cannon} exhibit good consistency with the results from our method, except that both sets of results also show systematic deviations in $T_{\rm eff}$ and [Fe/H] for dwarf stars ($T_{\rm eff}<4800$\,K), and $\log g$ for giant stars ($T_{\rm eff}<4300$\,K). These differences in $T_{\rm eff}$, $\log g$, and [Fe/H] mainly reflect the systematic differences in our training set, LAMOST DR5, and the reference catalogs. Direct comparisons of the LAMOST DR5, APOGEE--{\it Payne}, APOGEE--{\it ASPCAP}, and GALAH--{\it Cannon} stellar parameters for stars in common are presented in the Appendix (Fig.\,\ref{fig:label_cross_parameter}). It shows consistent patterns with those presented in Fig.\,\ref{fig:cross_parameter}, demonstrating that {\it CSNet} has merely inherited the systematic errors from the training sets.

\begin{figure*}
  \centering
  \includegraphics[width=\textwidth]{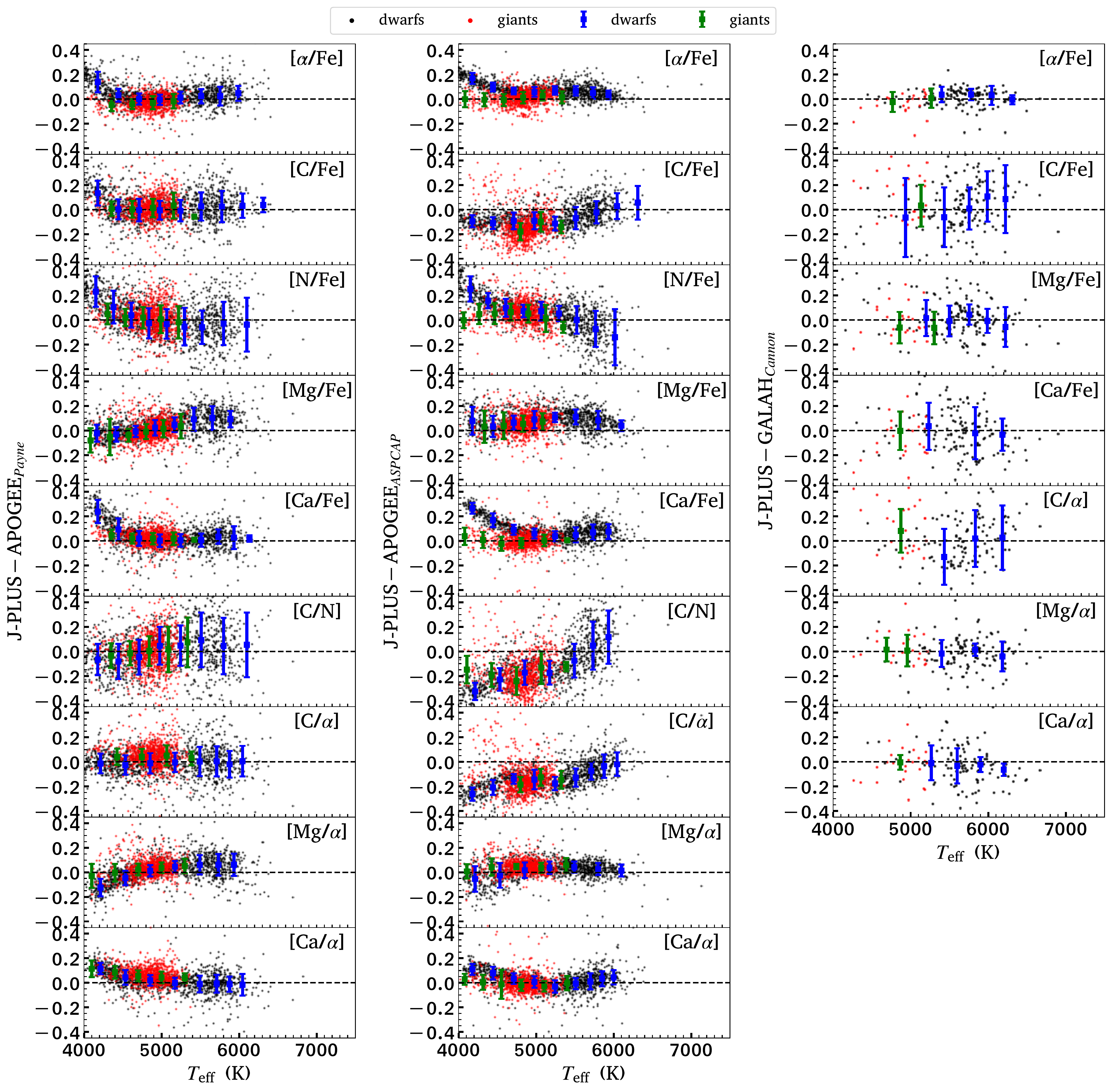}
  \caption{Similar to Fig.\,\ref{fig:lamost_abundance}, but for stars in common between the {\it CSNet} results and the reference catalogs.}
  \label{fig:cross_abundance}
\end{figure*}

Fig.\,\ref{fig:cross_abundance} shows the comparisons of elemental abundances between our estimates and that of APOGEE--{\it Payne}, APOGEE--{\it ASPCAP} and GALAH--{\it Cannon}. A close inspection suggests that there are systematic offsets and trends for [Mg/Fe] and [Ca/Fe] compared to the results from of APOGEE--{\it Payne} and APOGEE--{\it ASPCAP}. However, our [Mg/Fe] estimates are more consistent with that of GALAH--{\it Cannon}. This is because the [Mg/Fe] in the LAMOST--{\it DD-Payne} catalog, which is our training set, is derived using the GALAH--{\it Cannon} values as a reference (see Section\,\ref{LAMOST DR5}). Systematic deviations of the other elemental abundances (e.g., [$\alpha/$Fe], [C/Fe], [N/Fe], and [Ca/Fe]) are inherited from the differences among the reference catalogs, as consistent difference patterns exist in the training sets shown in the Appendix (Fig.\,\ref{fig:label_cross_abundance}).

\subsection{The J-PLUS DR1 catalog of stellar parameters and chemical abundances}

We select  4,387,568 stars (MAGABDUALOBJ\_CLASS\_STAR $\geq$ 0.6) by cross-matching J-PLUS DR1 with {\it Gaia} DR2, and determine their stellar parameters and chemical abundances using {\it CSNet}. The catalog is publicly available\footnote{\url{http://www.j-plus.es/ancillarydata/index}} . 
A description of the J-PLUS {\it CSNet} stellar parameters and chemical abundances catalog is provided in Table\,\ref{table:table2}. 

\begin{figure*}
  \centering
  \includegraphics[width=\textwidth]{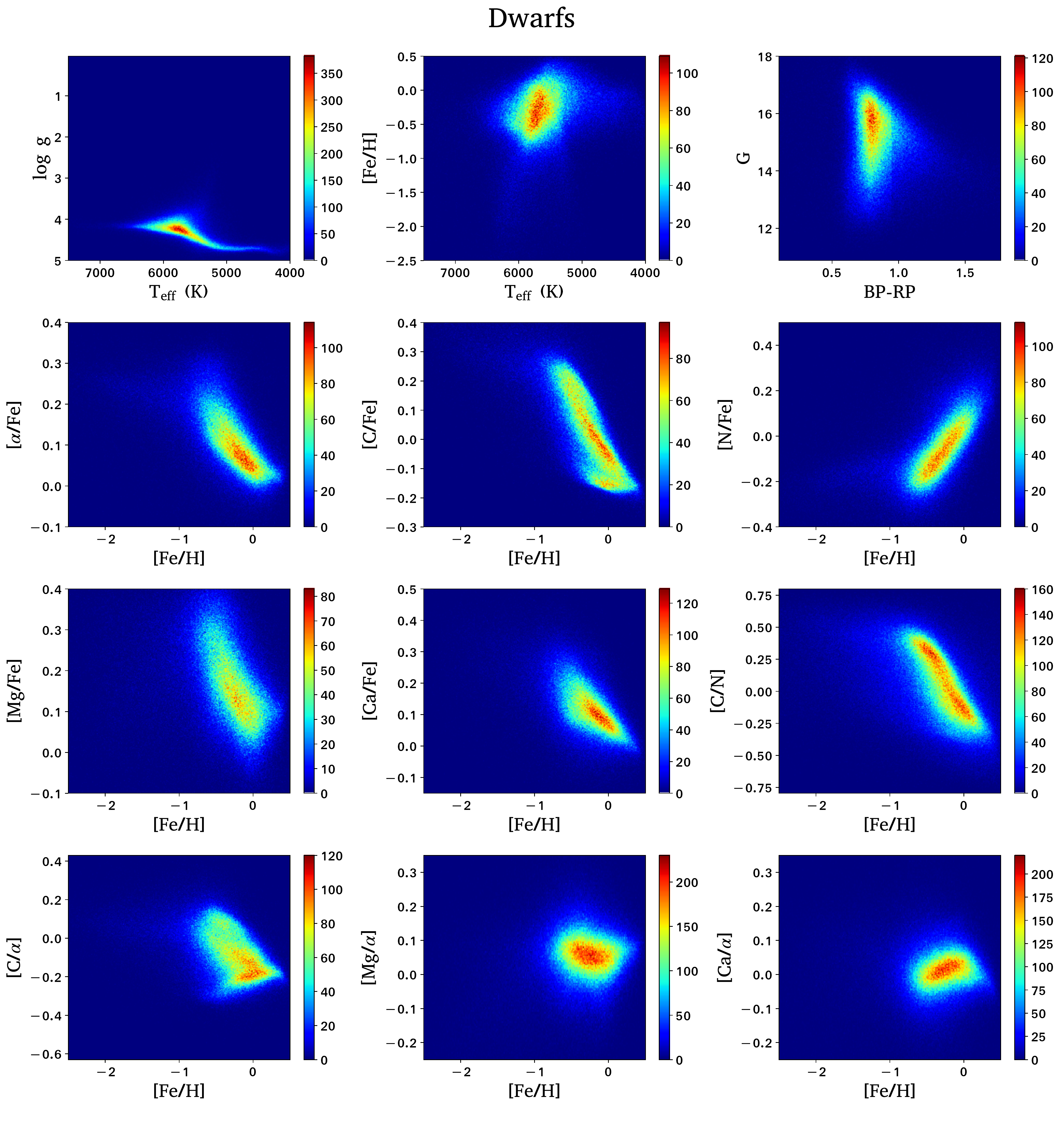}
  \caption{Density distributions of selected J-PLUS DR1 dwarf stars
  in the planes of $T_{\rm eff}$--$\log g$, $T_{\rm eff}$--[Fe/H], ($BP-RP$)--$G$, and different {\it CSNet} abundances with respect to [Fe/H], all color-coded by stellar number density. Only stars with reliable labels by the following criteria are used: (1) FLAGS = 0; (2) $0.063 < BP-RP < 1.786$; (3) $G<18$; (4) err (all J-PLUS filters) < 0.1 mag.}
  \label{fig:jplusgaia_d}
\end{figure*}

\begin{figure*}
  \centering
  \includegraphics[width=\textwidth]{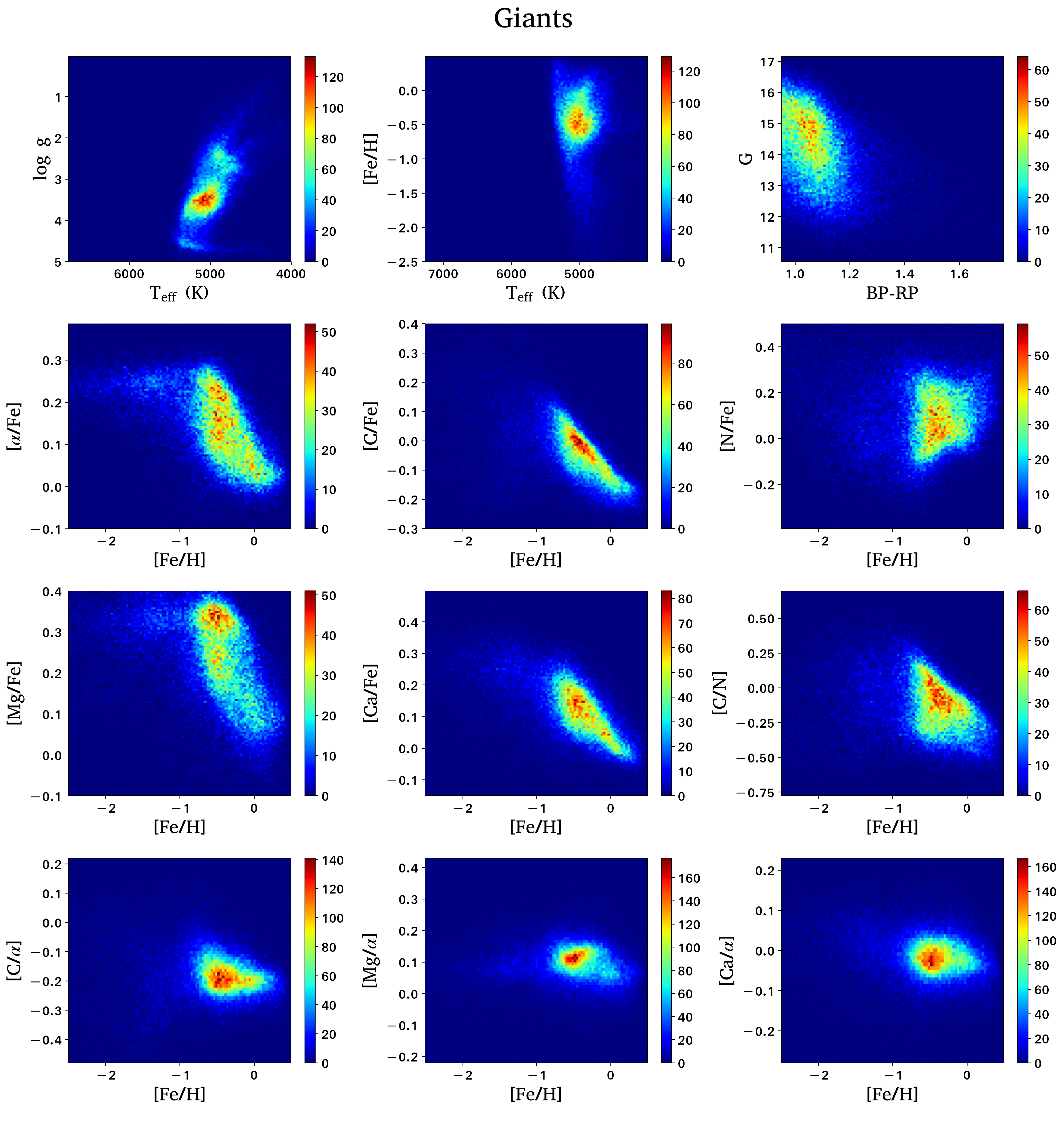}
  \caption{Similar to Fig.\,\ref{fig:jplusgaia_d}, but for selected J-PLUS DR1 giant stars.}
  \label{fig:jplusgaia_g}
\end{figure*}

\begin{figure*}
  \centering
  \includegraphics[width=\textwidth]{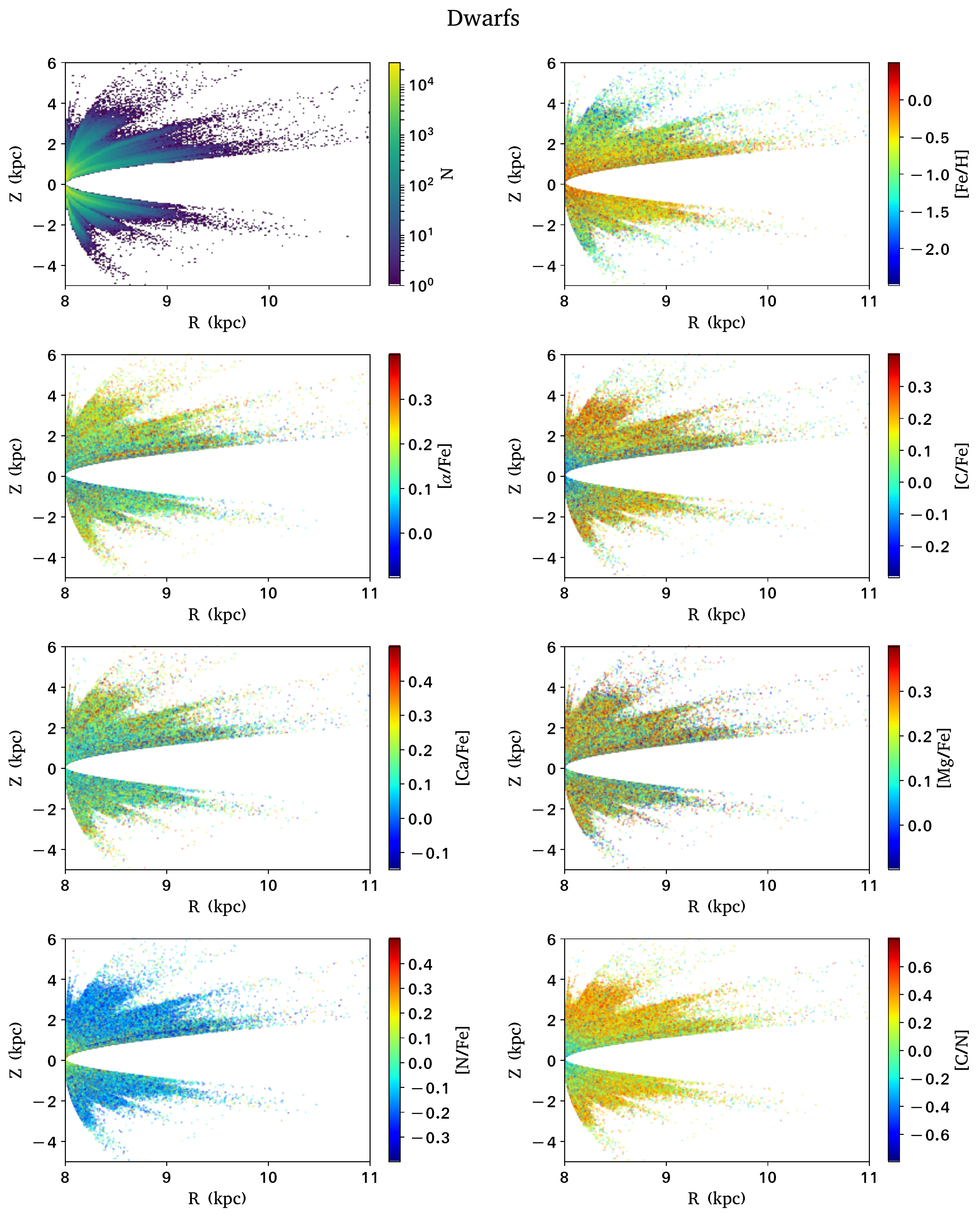}
  \caption{Distributions of number density and different {\it CSNet} abundances in the plane of R--Z for selected J-PLUS DR1 dwarf stars. Only stars with reliable labels by the following criteria are used: (1) FLAGS = 0; (2) $0.063 < BP-RP < 1.786$; (3) $G<18$; (4) err (all J-PLUS filters) < 0.1 mag.}
  \label{fig:rz_d}
\end{figure*}

\begin{figure*}
  \centering
  \includegraphics[width=\textwidth]{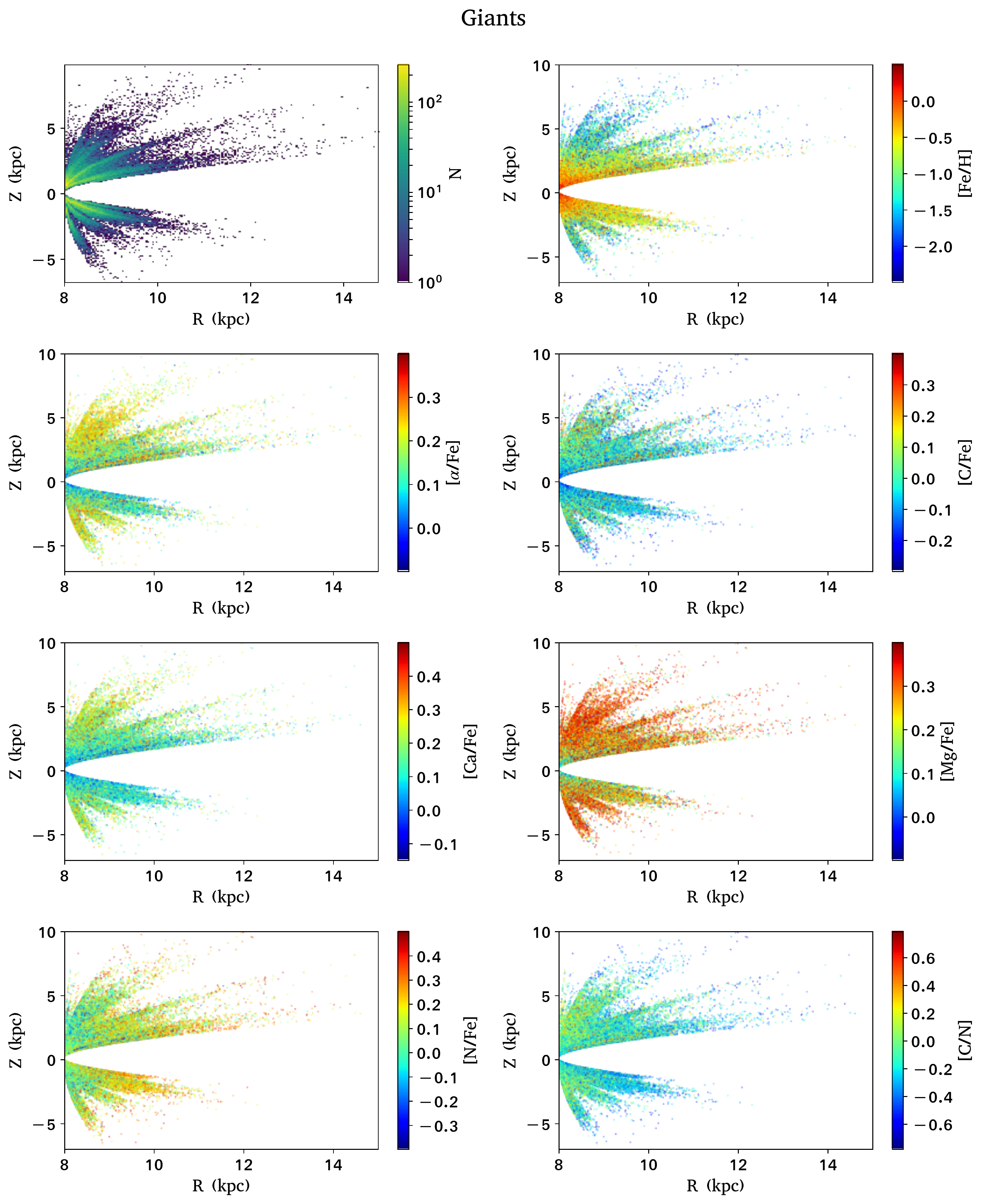}
  \caption{Similar to Fig.\,\ref{fig:rz_d}, but for selected J-PLUS DR1 giant stars.}
  \label{fig:rz_g}
\end{figure*}

Considering the photometric quality of J-PLUS DR1 and the limitations of {\it CSNet}, we recommend stellar labels for 2,343,597 stars with FLAGS $=$ 0 in all 12 J-PLUS filters and $0.063 < BP-RP < 1.786$. As discussed in Section\,\ref{Comparison}, $T_{\rm eff}$ for dwarf stars ($T_{\rm eff}<4800$\,K), and $\log g$ for giant stars ($T_{\rm eff}<4500$\,K) in our results should be used with caution, because they show non-negligible systematic errors related to the $T_{\rm eff}$ values. 
To avoid large label uncertainties caused by photometric errors, particularly for elemental abundances, 
we further select 0.61 million stars with $G < 18$ and magnitude errors in the 12 J-PLUS filters less than 0.1 mag, including 0.57 million dwarfs and 44,686 giants. Note that the giants and dwarfs are distinguished 
in the color-magnitude diagram ($BP-RP>0.95$ and $M_G<3.9$ for giants;  
$BP-RP\leq 0.95$ or $M_G\geq 3.9$ for dwarfs).
Fig.\,\ref{fig:jplusgaia_d} and \ref{fig:jplusgaia_g} show the stellar density distributions 
in the planes of $T_{\rm eff}$--$\log g$, $T_{\rm eff}$--[Fe/H], ($BP-RP$)--$G$, and different 
{\it CSNet} abundances  with respect to [Fe/H] for the 0.57 million dwarfs and 44,686 giants, respectively. 
Their distributions in the  $T_{\rm eff}$--$\log g$ diagrams are consistent with 
those in the color-magnitude diagrams.
The abundance trends are encouraging, and are consistent with literature results from LAMOST {\it DD--Payne} (Fig.\,\ref{fig:jplusgaia_d_train} and Fig.\,\ref{fig:jplusgaia_g_train}).
Fig.\,\ref{fig:rz_d} and \ref{fig:rz_g} show distributions of the number density and different {\it CSNet} abundances in the R--Z plane for the same sets of dwarfs and giants, respectively. Similarly, the abundance trends perform as expected. For example, values of [Fe/H] decrease and values of [Mg/Fe] and [C/N] increase with increasing distance from the Galactic plane. Detailed scientific investigations of the catalog, such as stellar populations, Galactic components and gradients based on these abundance results, will be presented in the future.

\begin{table*}
  \centering
  \caption{Description for the J-PLUS {\it CSNet} stellar parameters and chemical abundances catalog.\tnote{1}}
  \begin{threeparttable}
  \label{table:table3}
  \begin{tabular}{lll}
  \hline
   Col. &Field &    Description  \\
  \hline        
   1& specid &   J-PLUS source id \\
   2& gl &   Galactic longitude from the J-PLUS DR1 catalog (deg)\\
   3& gb &  Galactic latitude from the J-PLUS DR1 catalog (deg)  \\ 
   4& ra &   Right ascension from the J-PLUS DR1 catalog (J2000; deg)\\
   5& dec &  Declination from the J-PLUS DR1 catalog (J2000; deg)  \\
   6& MAG\_APER6 &  Photometry in 12 filters from the re-calibrated J-PLUS DR1 catalog   \\
   7& ERR\_APER6 &  Uncertainty in MAG\_APER6 (mag) \\
   8& FLAGS &   Inherited from SExtractor's FLAGS parameter  \\
   9& PHOT\_BP\_MEAN\_MAG &  $BP$-band photometry from {\it Gaia} DR2   \\
   10& PHOT\_RP\_MEAN\_MAG & $RP$-band photometry from {\it Gaia} DR2  \\
   11& PHOT\_G\_MEAN\_MAG &  $G$-band photometry from {\it Gaia} DR2  \\   
   12& $E(B-V)$ &  Interstellar extinction from the \citet{Schlegel1998} dust-reddening map \\
   7& $T_{\rm eff}$ &  Effective temperature (K)   \\
   8& $T_{\rm eff}$\_flag\tnote{1} &  A quality flag for $T_{\rm eff}$ \\
   9& $\log g$ &  Surface gravity  \\
   10& $\log g$\_flag\tnote{1} & A quality flag for $\log g$ \\
   11& {\rm [Fe/H]} & Iron to hydrogen abundance ratio  \\
   12& {\rm [Fe/H]}\_flag\tnote{1} &  A quality flag for {\rm [Fe/H]} \\
   13& {\rm [$\alpha$/Fe]} & $\alpha$-element to iron abundance ratio  \\
   14& {\rm [$\alpha$/Fe]}\_flag\tnote{1} & A quality flag for {\rm [$\alpha$/Fe]} \\
   15& {\rm [C/Fe]} & Carbon to iron abundance ratio   \\
   16& {\rm [C/Fe]}\_flag\tnote{1} & A quality flag for {\rm [C/Fe]}  \\
   17& {\rm [N/Fe]} & Nitrogen to iron abundance ratio  \\
   18& {\rm [N/Fe]}\_flag\tnote{1} & A quality flag for {\rm [N/Fe]} \\
   19& {\rm [Mg/Fe]} & Magnesium to iron abundance ratio  \\ 
   20& {\rm [Mg/Fe]}\_flag\tnote{1} & A quality flag for {\rm [Mg/Fe]}  \\
   21& {\rm [Ca/Fe]} & Calcium to iron abundance ratio  \\
   22& {\rm [Ca/Fe]}\_flag\tnote{1} & A quality flag for {\rm [Ca/Fe]}  \\
   \hline
  \end{tabular}
  \begin{tablenotes}
  \item[1] Flag = 0 (reliable) means that the object's $BP-RP$ color is within the effective $BP-RP$ range given in Table\,2, flag = 1 otherwise. 
  \end{tablenotes}
  \end{threeparttable}
\end{table*}

\section{Discussion}
\label{discussion} 

We combine the recalibrated J-PLUS DR1 and {\it Gaia} DR2 to construct 13 stellar colors. Then, the cost-sensitive neural-network based {\it CSNet} algorithm is designed and trained to map from the 13 colors to precise stellar labels. 
Thanks to the specially designed J-PLUS filters and Gaia BP/RP passbands, {\it CSNet} can not only determine the basic stellar
atmospheric parameters (effective temperature, $T_{\rm eff}$, surface gravity, $\log g$, and metallicity, [Fe/H]), but also deliver  [$\alpha$/Fe] and elemental abundances 
including [C/Fe], [N/Fe], [Mg/Fe], and [Ca/Fe]. 
This method performs well even if the training set has a sub-optimal distribution of stellar sample properties by increasing the error penalty for the rare subsets of stars. Our results show a high level of agreement with those from the testing samples. Comparisons with the APOGEE--{\it Payne}, APOGEE--{\it ASPCAP}, and GALAH--{\it Cannon} stars 
also show a good agreement, although some systematic discrepancies do exist, mainly caused by the 
systematic errors between the different surveys (see more details in Appendix\,\ref{LAMOST DR5 errors}). 
 
We also investigate the accuracy of the {\it CSNet} approach to derive the reddening, $E(B-V)$. Additional experiments show that {\it CSNet} is capable of estimating the $E(B-V)$ from stellar colors with a 1$\sigma$ uncertainty smaller than 0.02 mag. When {\it CSNet} is trained using the stellar colors without reddening correction, it achieves slightly lower accuracy for most stellar labels, suggesting that 
the J-PLUS filters should work well in regions of high extinction. 

Deep learning networks, including {\it CSNet}, require sufficient data with known labels to train and test a reliable model, which limits the coverage of the stellar labels that can be predicted. In the future, we plan to improve our method  to estimate [Fe/H] for very and extremely metal-poor stars. In addition, there are chemically peculiar stars (e.g., the carbon-enhanced metal-poor stars) with abundance ratios that lie outside the ranges we presently explore. We plan to remedy this limitation with the addition of such stars to the training/testing samples in the near future.

We point out that the J-PLUS and S-PLUS surveys are ongoing, and rapidly growing in their coverage of the Northern and Southern Hemispheres, respectively.  Application of {\it CSnet}, and its planned refinements, to both of these datasets will eventually be able to provide similar results as presented in this paper for hundreds of millions of stars over much of the sky.

\section{Summary}
\label{summary} 

By combining photometric data from the recalibrated J-PLUS DR1, {\it Gaia} DR2, and spectroscopic labels from LAMOST, we design and train a cost-sensitive neural network, {\it CSNet}, to learn the non-linear mapping from stellar colors to their labels. 
Special attention is paid to the minority populations in the label space by assigning different weights according to their density distributions. 
Thanks to the specially designed J-PLUS narrow-band filters, 
{\it CSNet} can not only determine the basic stellar
atmospheric parameters (effective temperature, $T_{\rm eff}$, surface gravity, $\log g$, and metallicity, [Fe/H]), but also deliver  [$\alpha$/Fe] and elemental abundances 
including [C/Fe], [N/Fe], [Mg/Fe], and [Ca/Fe]. 
We have achieved precisions of $\delta\, T_{\rm eff}\sim$ 55\,K, $\delta \log g\sim 0.15$\,dex, and $\delta$ [Fe/H] $\sim$ 0.07\,dex, respectively. The uncertainties of the abundance estimates for [$\alpha$/Fe] and the four individual elements are on the order of $\delta$\, $\sim$ 0.04--0.08\,dex. We compare our parameter and abundance estimates with those from other spectroscopic catalogs such as APOGEE and GALAH, finding an overall good agreement.
Applying our method to J-PLUS DR1, we have obtained the aforementioned parameters for over two million stars, 
providing a powerful data set for chemo-dynamical analyses of the Milky Way. The catalog of the estimated parameters is publicly accessible. 
Our results also demonstrate the potential of well-designed and high-quality photometric data in determinations 
of stellar parameters as well as for individual elemental abundances.

\begin{acknowledgements}

The authors thank Marwan Gebran for his detailed reading and suggestions that improved the clarity of our presentation. We acknowledge David Sobral for a careful reading of the manuscript. This work is supported by the National Natural Science Foundation of China through the projects NSFC 12173007, 11603002, National Key Basic R \& D Program of China via 2019YFA0405500, 
and Beijing Normal University grant No. 310232102. 
T.C.B. acknowledges partial support from grant PHY 14-30152,
Physics Frontier Center/JINA Center for the Evolution
of the Elements (JINA-CEE), awarded by the US National
Science Foundation. His participation in this work was initiated
by conversations that took place during a visit to China
in 2019, supported by a PIFI Distinguished Scientist award
from the Chinese Academy of Science. C.A.G. acknowledges financial support from the CAPES Brazilian Funding Agency. We acknowledge the science research grants from the China Manned Space Project with NO. CMS-CSST-2021-A08 and CMS-CSST-2021-A09. J. V. acknowledges the technical members of the UPAD for their invaluable work: Juan Castillo, Tamara Civera, Javier Hernández, Ángel López, Alberto Moreno, and David Muniesa. JAFO acknowledges the financial support from the Spanish Ministry of Science and Innovation and the European Union -- NextGenerationEU through the Recovery and Resilience Facility project ICTS-MRR-2021-03-CEFCA.Based on observations made with the JAST/T80 telescope at the Observatorio Astrof{\'i}sico de Javalambre (OAJ), in Teruel, owned, managed and operated by the Centro de Estudios de F{\'i}sica del Cosmos de Arag{\'o}n. We acknowledge the OAJ Data Processing and Archiving Unit (UPAD) for reducing and calibrating the OAJ data used in this work. Funding for the J-PLUS Project has been provided by the Governments of Spain and Arag{\'o}n through the Fondo de Inversiones de Teruel; the Arag{\'o}n Government through the Reseach Groups E96, E103, and E16\_17R; the Spanish Ministry of Science, Innovation and Universities (MCIU/AEI/FEDER, UE) with grants PGC2018-097585-B-C21 and PGC2018-097585-B-C22, the Spanish Ministry of Economy and Competitiveness (MINECO) under AYA2015- 66211-C2-1-P, AYA2015-66211-C2-2, AYA2012-30789, and ICTS-2009-14; and European FEDER funding (FCDD10-4E-867, FCDD13-4E-2685). The Brazilian agencies FINEP, FAPESP, and the National Observatory of Brazil have also contributed to this project.
Guoshoujing Telescope (the Large Sky Area Multi-Object Fiber Spectroscopic Telescope LAMOST) is a National Major Scientific Project built by the Chinese Academy of Sciences. Funding for the project has been provided by the National Development and Reform Commission. LAMOST is operated and managed by the National Astronomical Observatories, Chinese Academy of Sciences.
This work has made use of data from the European Space Agency (ESA) mission {\it Gaia} (https://www.cosmos.esa.int/gaia), processed by the {\it Gaia} Data Processing and Analysis
Consortium (DPAC, https://www.cosmos.esa.int/
web/gaia/dpac/ consortium). Funding for the DPAC has been provided by national institutions, in particular the institutions participating in the {\it Gaia} Multilateral Agreement. 

\end{acknowledgements}

\bibliographystyle{aa}

\bibliography{Bibliography}

\begin{thebibliography}{64}
\expandafter\ifx\csname natexlab\endcsname\relax\def\natexlab#1{#1}\fi

\bibitem[Alvarez \& Plez(1998)]{Alvarez1998} Alvarez, R. \& Plez, B.\ 1998, \aap, 330, 1109
\bibitem[Almeida-Fernandes et al.(2021)]{Almeida-Fernandes2021} Almeida-Fernandes, F., Sampedro, L., Herpich, F.~R., et al.\ 2021, arXiv:2104.00020
\bibitem[{\'A}rnad{\'o}ttir et al.(2010)]{Arnadottir2010} {\'A}rnad{\'o}ttir, A.~S., Feltzing, S., \& Lundstr{\"o}m, I.\ 2010, \aap, 521, A40. doi:10.1051/0004-6361/200913544
\bibitem[Bai et al.(2019)]{Bai2019} Bai, Y., Liu, J., Bai, Z., et al.\ 2019, \aj, 158, 93. doi:10.3847/1538-3881/ab3048
\bibitem[Bailer-Jones(2011)]{Bailer2011} Bailer-Jones, C.~A.~L.\ 2011, \mnras, 411, 435. doi:10.1111/j.1365-2966.2010.17699.x
\bibitem[Bailer-Jones(2002)]{Bailer2002} Bailer-Jones, C.~A.~L.\ 2002, \apss, 280, 21. doi:10.1023/A:1015527705755
\bibitem[Benitez et al.(2014)]{Benitez2014} Benitez, N., Dupke, R., Moles, M., et al.\ 2014, arXiv:1403.5237
\bibitem[Buder et al.(2018)]{Buder2018} Buder, S., Asplund, M., Duong, L., et al.\ 2018, \mnras, 478, 4513. doi:10.1093/mnras/sty1281
\bibitem[Cardamone et al.(2010)]{Cardamone2010} Cardamone, C.~N., van Dokkum, P.~G., Urry, C.~M., et al.\ 2010, \apjs, 189, 270. doi:10.1088/0067-0049/189/2/270
\bibitem[Casagrande et al.(2019)]{Casagrande2019} Casagrande, L., Wolf, C., Mackey, A.~D., et al.\ 2019, \mnras, 482, 2770. doi:10.1093/mnras/sty2878
\bibitem[Cenarro et al.(2014)]{Cenarro2014} Cenarro, A.~J., Moles, M., Mar{\'\i}n-Franch, A., et al.\ 2014, \procspie, 9149, 91491I. doi:10.1117/12.2055455
\bibitem[Cenarro et al.(2019)]{Cenarro2019} Cenarro, A.~J., Moles, M., Crist{\'o}bal-Hornillos, D., et al.\ 2019, \aap, 622, A176. doi:10.1051/0004-6361/201833036
\bibitem[Chao et al.(2013)]{Chao2013} Chao, W.-L, Liu, J.-Z., \& Ding, J.-J.\ 2013, Pattern Recognition, 46, 628
\bibitem[Chiti et al.(2020)]{Chiti2020} Chiti, A., Frebel, A., Jerjen, H., et al.\ 2020, \apj, 891, 8. doi:10.3847/1538-4357/ab6d72
\bibitem[Chiti et al.(2021)]{Chiti2021} Chiti, A., Frebel, A., Mardini, M.~K., et al.\ 2021, \apjs, 254, 31. doi:10.3847/1538-4365/abf73d
\bibitem[Cui et al.(2012)]{Cui2012} Cui, X.-Q., Zhao, Y.-H., Chu, Y.-Q., et al.\ 2012, Research in Astronomy and Astrophysics, 12, 1197. doi:10.1088/1674-4527/12/9/003
\bibitem[Deng et al.(2012)]{Deng2012} Deng, L.-C., Newberg, H.~J., Liu, C., et al.\ 2012, Research in Astronomy and Astrophysics, 12, 735. doi:10.1088/1674-4527/12/7/003
\bibitem[De Silva et al.(2015)]{De_Silva2015} De Silva, G.~M., Freeman, K.~C., Bland-Hawthorn, J., et al.\ 2015, \mnras, 449, 2604. doi:10.1093/mnras/stv327
\bibitem[Evans et al.(2018)]{Evans2018} Evans, D.~W., Riello, M., De Angeli, F., et al.\ 2018, \aap, 616, A4. doi:10.1051/0004-6361/201832756
\bibitem[Gaia Collaboration et al.(2016)]{GaiaDR22016} Gaia Collaboration, Prusti, T., de Bruijne, J.~H.~J., et al.\ 2016, \aap, 595, A1. doi:10.1051/0004-6361/201629272
\bibitem[Gaia Collaboration et al.(2018)]{Brown2018} Gaia Collaboration, Brown, A.~G.~A., Vallenari, A., et al.\ 2018, \aap, 616, A1. doi:10.1051/0004-6361/201833051
\bibitem[Galarza et al.(2021)]{Andres2021} Galarza, C.~A., Daflon, S., Placco, V.~M., et al.\ 2021, arXiv:2109.11600
\bibitem[Gao et al.(2016)]{Gao2016} Gao, X.-Y, Chen, Z.-Y, Tang, S., et al.\ 2016, Neurocomputing, 173, 1927
\bibitem[Garc{\'\i}a P{\'e}rez et al.(2016)]{Perez2016} Garc{\'\i}a P{\'e}rez, A.~E., Allende Prieto, C., Holtzman, J.~A., et al.\ 2016, \aj, 151, 144. doi:10.3847/0004-6256/151/6/144
\bibitem[Green et al.(2021)]{Green2021} Green, G.~M., Rix, H.~W., Tschesche, J., et al.\ 2021, \apj, 907, 57
\bibitem[Holtzman et al.(2018)]{Holtzman2018} Holtzman, J.~A., Hasselquist, S., Shetrone, M., et al.\ 2018, \aj, 156, 125. doi:10.3847/1538-3881/aad4f9
\bibitem[Huang \& Yuan (2021)]{HuangYuan2021} Huang, B.-W. \& Yuan, H.-B., \ 2021, \apjs, submitted
\bibitem[Huang et al.(2019)]{Huang2019} Huang, Y., Chen, B.-Q., Yuan, H.-B., et al.\ 2019, \apjs, 243, 7. doi:10.3847/1538-4365/ab1f72
\bibitem[Huang et al.(2021a)]{Huang2021a} Huang, Y., Beers, T.~C., Wolf, C., et al.\ 2021a, arXiv:2104.14154
\bibitem[Huang et al.(2021b)]{Huang2021b} Huang, Y., Yuan, H., Li, C., et al.\ 2021b, \apj, 907, 68. doi:10.3847/1538-4357/abca37
\bibitem[Ilbert et al.(2009)]{Ilbert2009} Ilbert, O., Capak, P., Salvato, M., et al.\ 2009, \apj, 690, 1236. doi:10.1088/0004-637X/690/2/1236
\bibitem[Jones et al.(2014)]{Jones2014} Jones, O.~C., Kemper, F., Srinivasan, S., et al.\ 2014, \mnras, 440, 631. doi:10.1093/mnras/stu286
\bibitem[L{\'o}pez-Sanjuan et al.(2019)]{Sanjuan2019} L{\'o}pez-Sanjuan, C., Varela, J., Crist{\'o}bal-Hornillos, D., et al.\ 2019, \aap, 631, A119. doi:10.1051/0004-6361/201936405
\bibitem[L{\'o}pez-Sanjuan et al.(2021)]{Sanjuan2021} L{\'o}pez-Sanjuan, C., Yuan, H., V{\'a}zquez Rami{\'o}, H., et al.\ 2021, \aap, 654, A61. doi:10.1051/0004-6361/202140444
\bibitem[Luo et al.(2012)]{LAMOSTDR5Catalog} Luo, A.-L., Zhang, H.-T., Zhao, Y.-H., et al.\ 2012, Research in Astronomy and Astrophysics, 12, 1243. doi:10.1088/1674-4527/12/9/004
\bibitem[Luo et al.(2015)]{Luo2015} Luo, A.-L., Zhao, Y.-H., Zhao, G., et al.\ 2015, Research in Astronomy and Astrophysics, 15, 1095. doi:10.1088/1674-4527/15/8/002
\bibitem[Kingma \& Ba(2014)]{Adam} Kingma, D.~P. \& Ba, J.\ 2014, arXiv:1412.6980
\bibitem[Ksoll et al.(2020)]{Ksoll2020} Ksoll, V.~F., Ardizzone, L., Klessen, R., et al.\ 2020, \mnras, 499, 5447. doi:10.1093/mnras/staa2931
\bibitem[Liu et al.(2006)]{Liu2006} Liu, X., Wu, J., \& Zhou, Z.\ 2006, Sixth International Conference on Data Mining (ICDM'06), 965
\bibitem[Liu et al.(2014)]{Liu2014} Liu, X.-W., Yuan, H.-B., Huo, Z.-Y., et al.\ 2014, Setting the scene for Gaia and LAMOST, 298, 310. doi:10.1017/S1743921313006510
\bibitem[Majewski et al.(2017)]{Majewski2017} Majewski, S.~R., Schiavon, R.~P., Frinchaboy, P.~M., et al.\ 2017, \aj, 154, 94. doi:10.3847/1538-3881/aa784d
\bibitem[Marin-Franch et al.(2015)]{Marin2015} Marin-Franch, A., Taylor, K., Cenarro, J., et al.\ 2015, IAU General Assembly
\bibitem[Mendes de Oliveira et al.(2019)]{Mendes2019} Mendes de Oliveira, C., Ribeiro, T., Schoenell, W., et al.\ 2019, \mnras, 489, 241. doi:10.1093/mnras/stz1985
\bibitem[Miller et al.(2015)]{Miller2015} Miller, A.~A., Bloom, J.~S., Richards, J.~W., et al.\ 2015, \apj, 798, 122. doi:10.1088/0004-637X/798/2/122
\bibitem[Moles et al.(2008)]{Moles2008} Moles, M., Ben{\'\i}tez, N., Aguerri, J.~A.~L., et al.\ 2008, \aj, 136, 1325. doi:10.1088/0004-6256/136/3/1325
\bibitem[Nekooeimehr \& Susana(2015)]{Nekooeimehr2015} Nekooeimehr, I. \& Susana K.\ 2015, Expert Systems with Applications, 46
\bibitem[Ness et al.(2015)]{Ness2015} Ness, M., Hogg, D.~W., Rix, H.-W., et al.\ 2015, \apj, 808, 16. doi:10.1088/0004-637X/808/1/16
\bibitem[Niu et al.(2021a)]{Niu2021a} Niu, Z., Yuan, H., \& Liu, J.\ 2021a, \apj, 909, 48. doi:10.3847/1538-4357/abdbac
\bibitem[Niu et al.(2021b)]{Niu2021b} Niu, Z., Yuan, H., \& Liu, J.\ 2021b, \apjl, 908, L14. doi:10.3847/2041-8213/abe1c2

\bibitem[Onan \& Aytu\v{g}(2019)]{Onan2019} Onan, \& Aytu\v{g}.\ 2019, Scientific Programming, 2019, 1
\bibitem[Onken et al.(2019)]{Onken2019} Onken, C.~A., Wolf, C., Bessell, M.~S., et al.\ 2019, \pasa, 36, e033. doi:10.1017/pasa.2019.27

\bibitem[Oommen et al.(2011)]{Oommen2011} Oommen, Thomas, Baise, et al.\ 2011, Mathematical Geosciences, 43, 99
\bibitem[P{\'e}rez-Gonz{\'a}lez et al.(2013)]{PerezGonzalez2013} P{\'e}rez-Gonz{\'a}lez, P.~G., Cava, A., Barro, G., et al.\ 2013, \apj, 762, 46. doi:10.1088/0004-637X/762/1/46
\bibitem[Postman et al.(2012)]{Postman2012} Postman, M., Coe, D., Ben{\'\i}tez, N., et al.\ 2012, \apjs, 199, 25. doi:10.1088/0067-0049/199/2/25
\bibitem[Rossi et al.(2005)]{Rossi2005} Rossi, S., Beers, T.~C., Sneden, C., et al.\ 2005, \aj, 130, 2804. doi:10.1086/497164
\bibitem[Schlegel et al.(1998)]{Schlegel1998} Schlegel, D.~J., Finkbeiner, D.~P., \& Davis, M.\ 1998, \apj, 500, 525. doi:10.1086/305772
\bibitem[Ting et al.(2019)]{Ting2019} Ting, Y.-S., Conroy, C., Rix, H.-W., et al.\ 2019, \apj, 879, 69. doi:10.3847/1538-4357/ab2331
\bibitem[Wang et al.(2021a)]{2021arXiv210612787W} Wang, C., Bai, Y., L{\'o}pez-Sanjuan, C., et al.\ 2021a, arXiv:2106.12787
\bibitem[Wang et al.(2021b)]{Wang2021b} Wang, C., Bai, Y., Yuan, H, et al.\ 2021b, \aap, to be submitted
\bibitem[Whitten et al.(2019)]{Whitten2019} Whitten, D.~D., Placco, V.~M., Beers, T.~C., et al.\ 2019, \aap, 622, A182. doi:10.1051/0004-6361/201833368
\bibitem[Whitten et al.(2021)]{Whitten2021} Whitten, D.~D., Placco, V.~M., Beers, T.~C., et al.\ 2021, \apj, 912, 147. doi:10.3847/1538-4357/abee7e
\bibitem[Wolf et al.(2003)]{Wolf2003} Wolf, C., Meisenheimer, K., Rix, H.-W., et al.\ 2003, \aap, 401, 73. doi:10.1051/0004-6361:20021513
\bibitem[Wolf et al.(2018)]{Wolf2018} Wolf, C., Onken, C.~A., Luvaul, L.~C., et al.\ 2018, \pasa, 35, e010. doi:10.1017/pasa.2018.5
\bibitem[Wu et al.(2011)]{Wu2011} Wu, Y., Luo, A.-L., Li, H.-N., et al.\ 2011, Research in Astronomy and Astrophysics, 11, 924. doi:10.1088/1674-4527/11/8/006
\bibitem[Xiang et al.(2019)]{Xiang2019} Xiang, M., Ting, Y.-S., Rix, H.-W., et al.\ 2019, \apjs, 245, 34. doi:10.3847/1538-4365/ab5364
\bibitem[Xu et al. (2021)]{Xu2021} Xu, S., Yuan, H.-B., Niu, Z.-X., et al.\ 2021, \apjs, submitted 
\bibitem[Yang et al.(2021)]{Yang2021} Yang, L., Yuan, H., Zhang, R., et al.\ 2021, \apjl, 908, L24. doi:10.3847/2041-8213/abdbae
\bibitem[Yanny et al.(2009)]{Yanny2009} Yanny, B., Rockosi, C., Newberg, H.~J., et al.\ 2009, \aj, 137, 4377. doi:10.1088/0004-6256/137/5/4377
\bibitem[York et al.(2000)]{York2000} York, D.~G., Adelman, J., Anderson, J.~E., et al.\ 2000, \aj, 120, 1579. doi:10.1086/301513
\bibitem[Yuan et al.(2013)]{Yuan2013} Yuan, H.~B., Liu, X.~W., \& Xiang, M.~S.\ 2013, \mnras, 430, 2188. doi:10.1093/mnras/stt039
\bibitem[Yuan et al.(2015a)]{Yuan2015a} Yuan, H., Liu, X., Xiang, M., et al.\ 2015a, \apj, 799, 133. doi:10.1088/0004-637X/799/2/133
\bibitem[Yuan et al.(2015b)]{Yuan2015b} Yuan, H., Liu, X., Xiang, M., et al.\ 2015b, \apj, 799, 134. doi:10.1088/0004-637X/799/2/134
\bibitem[Yuan et al.(2015c)]{Yuan2015c} Yuan, H., Liu, X., Xiang, M., et al.\ 2015c, \apj, 803, 13. doi:10.1088/0004-637X/803/1/13
\bibitem[Zakaryazad \& Duman(2016)]{Zakaryazad2016} Zakaryazad, A., \& Duman, E.\ 2016, Neurocomputing, 175, 121
\bibitem[Zhang et al.(2019)]{938720Giants} Zhang, X., Zhao, G., Yang, C.~Q., et al.\ 2019, \pasp, 131, 094202. doi:10.1088/1538-3873/ab2687
\bibitem[Zhang et al.(2021)]{Zhang2021} Zhang, R., Yuan, H., Liu, X., et al.\ 2021, arXiv:2109.06390
\bibitem[Zheng et al.(2018)]{Zheng2018} Zheng, J., Zhao, G., Wang, W., et al.\ 2018, Research in Astronomy and Astrophysics, 18, 147. doi:10.1088/1674-4527/18/12/147
\bibitem[Zhu et al.(2017)]{Zhu2017} Zhu, T.-F, Lin, Y.-P., \& Liu, Y.-H. \ 2017, Pattern Recognition, 72, 327
\bibitem[Zong et al.(2013)]{Zong2013} Zong, W.-W, Huang, G.-B., \& Chen, Y.-Q.\ 2013, Neurocomputing, 101, 229



\end{thebibliography}

\begin{appendix}

\section{Correlation analyses between J-PLUS colors and stellar labels }\label{color and label}
The core idea of {\it CSNet} is to construct a mapping from J-PLUS colors to stellar labels. This leaves the question as to whether our results are from specific sensitive J-PLUS colors or just from correlations among the stellar labels themselves. 

\begin{figure*}
  \centering
  \includegraphics[width=\textwidth]{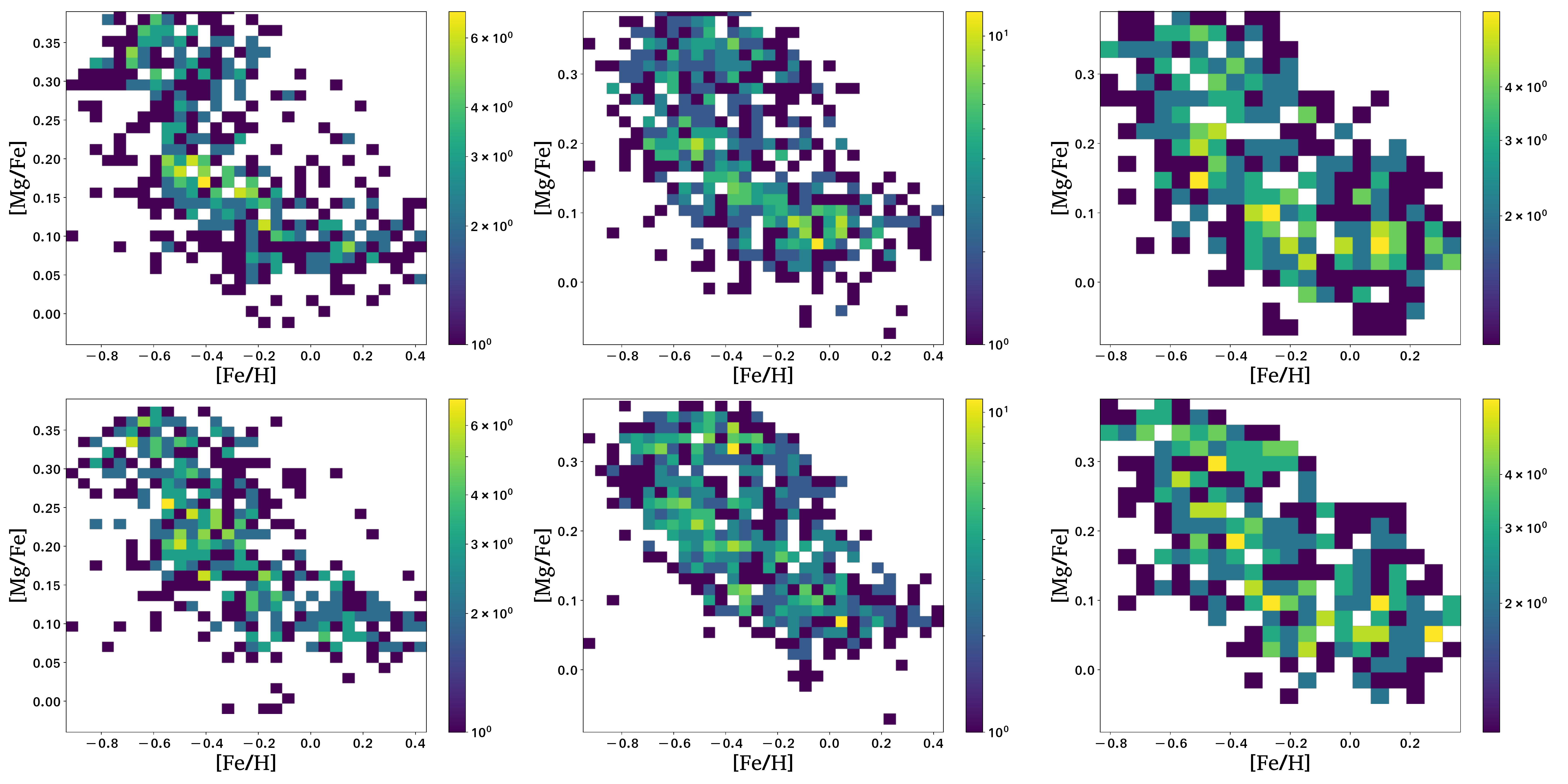}
  \caption{Stellar number density distributions in the plane of [Mg/Fe] -- [Fe/H] from the LAMOST catalog ({\it top} panels) and the {\it CSNet} results ({\it bottom} panels) for the training/testing sample giant stars, all color-coded by stellar number density. Different columns are for stars of different 
  $BP-RP$ ranges. From left to right these are [0.95, 1.05], [1.05, 1.15], and [1.15, 1.40], respectively.}
  \label{fig:mgfe_giant}
\end{figure*}

\begin{figure*}
  \centering
  \includegraphics[width=\textwidth]{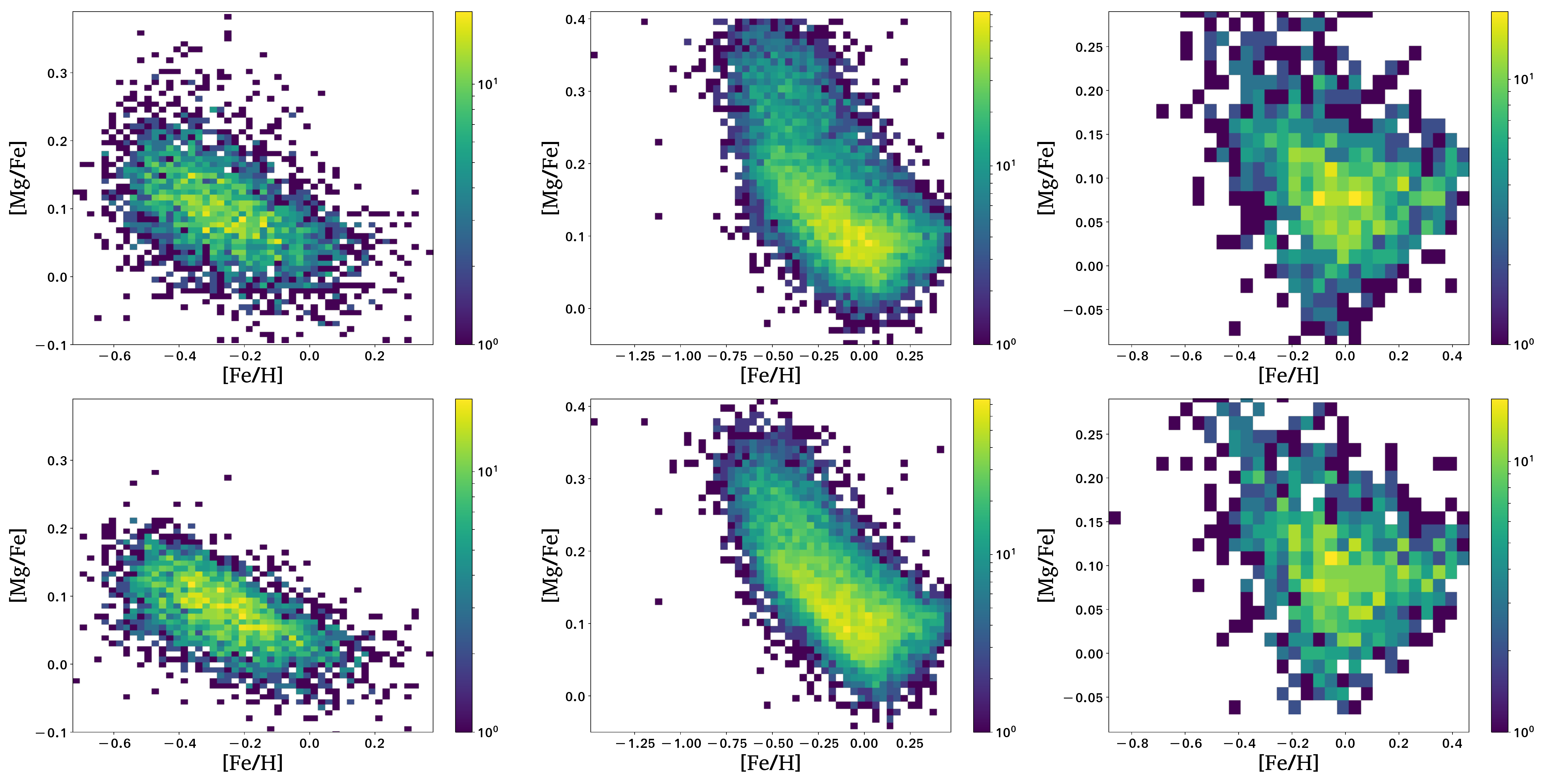}
  \caption{Similar to Fig.\,\ref{fig:mgfe_giant}, but for the dwarf stars. The $BP-RP$ ranges are [0.25, 0.70] ({\it left} column), [0.70, 1.00] ({\it middle} column) and [1.00, 1.50] ({\it right} column), respectively.}
  \label{fig:mgfe_dwarf}
\end{figure*}

\begin{figure*}
  \centering
  \includegraphics[width=\textwidth]{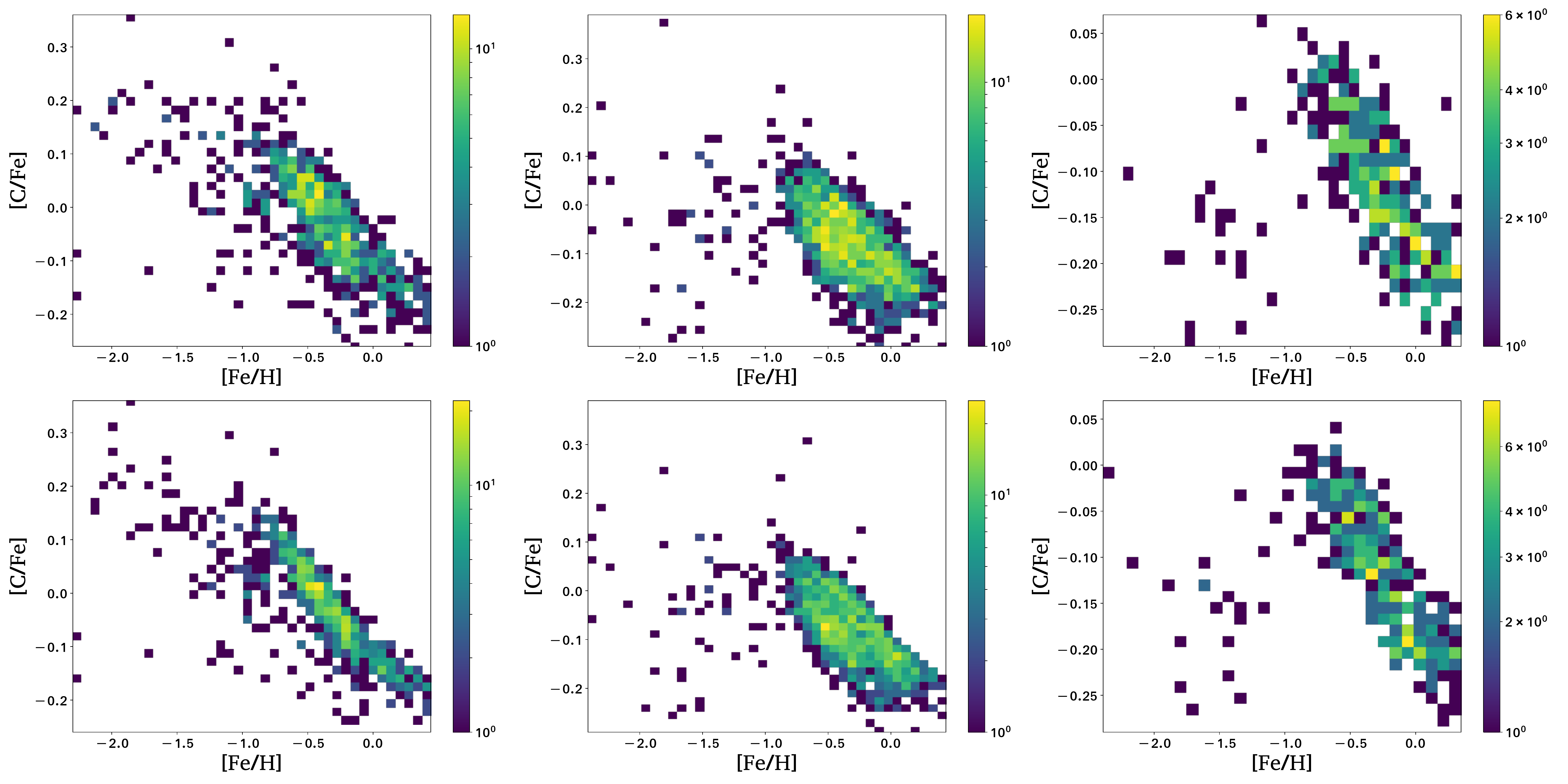}
  \caption{Stellar number density distributions in the plane of [C/Fe] -- [Fe/H] from the LAMOST {\it DD--Payne} ({\it top} panels) and the {\it CSNet} results ({\it bottom} panels) for the training/testing sample giant stars, all color-coded by  stellar number density. Different columns are for stars of different 
  $BP-RP$ ranges. From left to right these are [0.95, 1.05],  [1.05, 1.20], and [1.20, 1.60], respectively.}
  \label{fig:cfe_giant}
\end{figure*}

\begin{figure*}
  \centering
  \includegraphics[width=\textwidth]{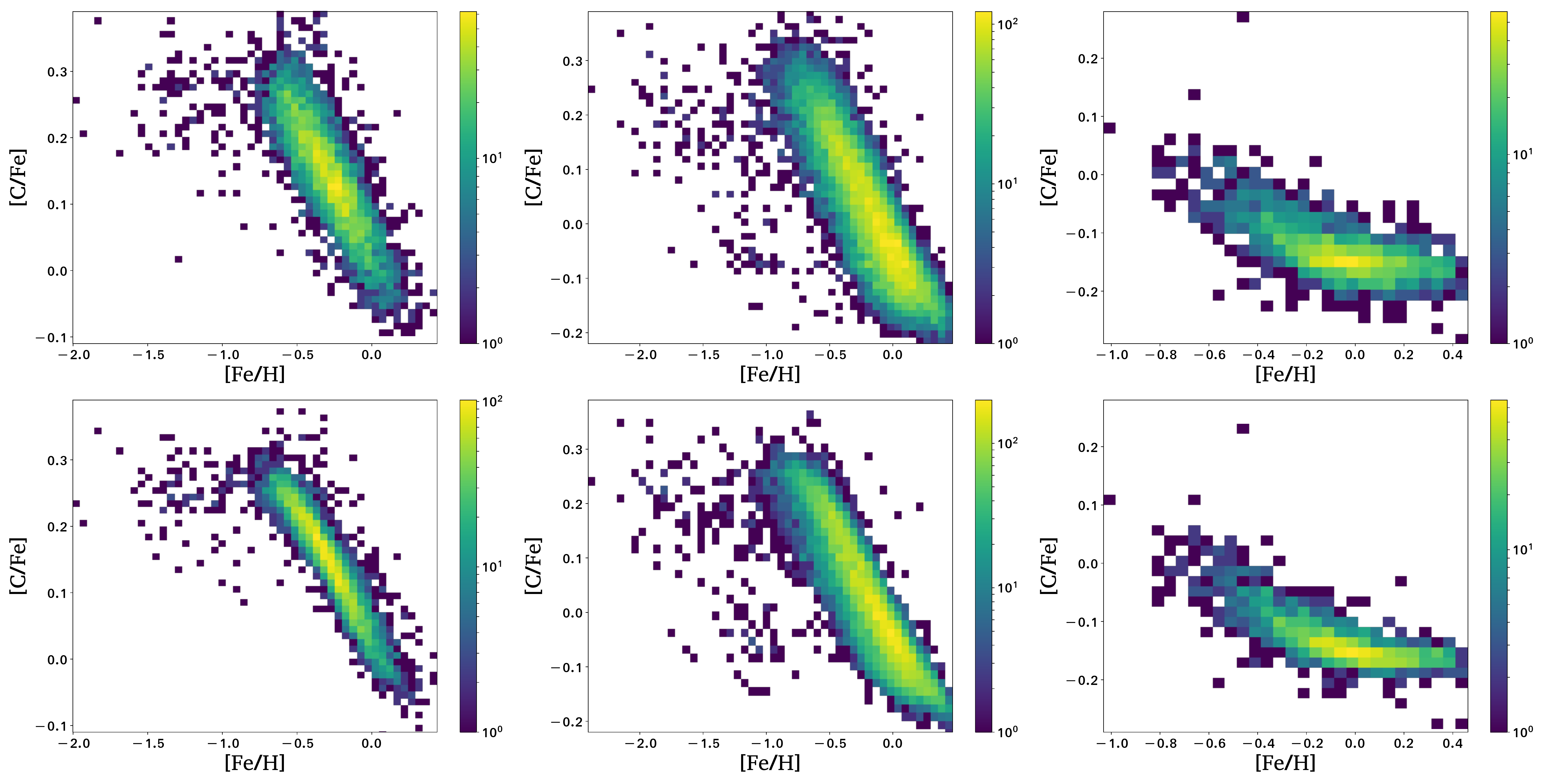}
  \caption{Similar to Fig.\,\ref{fig:cfe_giant}, but for the dwarf stars. The $BP-RP$ ranges are [0.25, 0.70] ({\it left} column), [0.70, 1.00] ({\it middle} column) and [1.00, 1.50] ({\it right} column), respectively.}
  \label{fig:cfe_dwarf}
\end{figure*}

\begin{figure*}
  \centering
  \includegraphics[width=\textwidth]{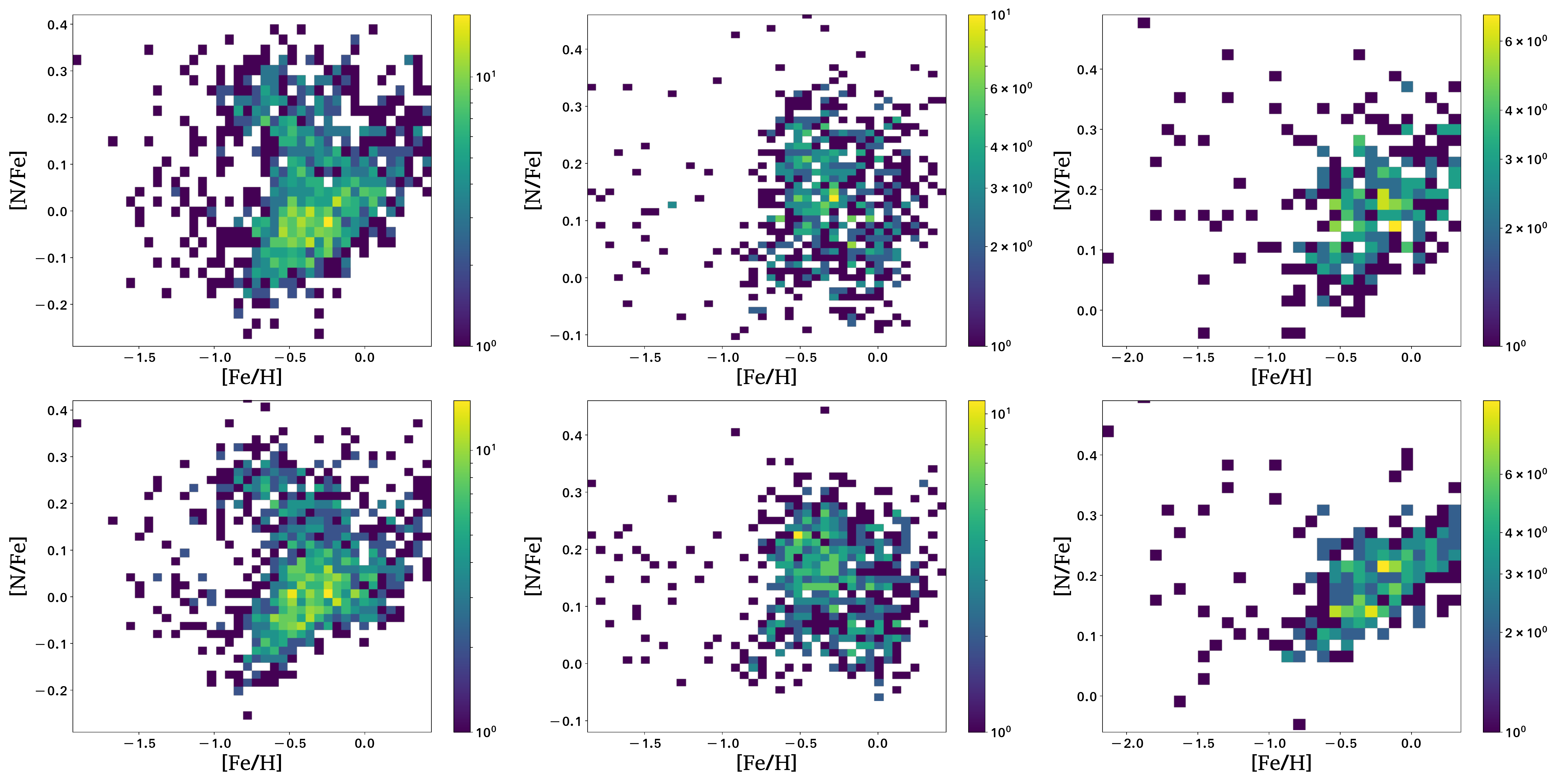}
  \caption{Stellar number density distributions in the plane of [N/Fe] -- [Fe/H] from the LAMOST {\it DD--Payne} ({\it top} panels) and the {\it CSNet} results ({\it bottom} panels) for the training/testing sample giant stars, all color-coded by stellar number density. Different columns are for stars of different 
  $BP-RP$ ranges. From left to right these are [0.95, 1.10], [1.10, 1.20], and [1.20, 1.60], respectively.}
  \label{fig:nfe_giant}
\end{figure*}

\begin{figure*}
  \centering
  \includegraphics[width=\textwidth]{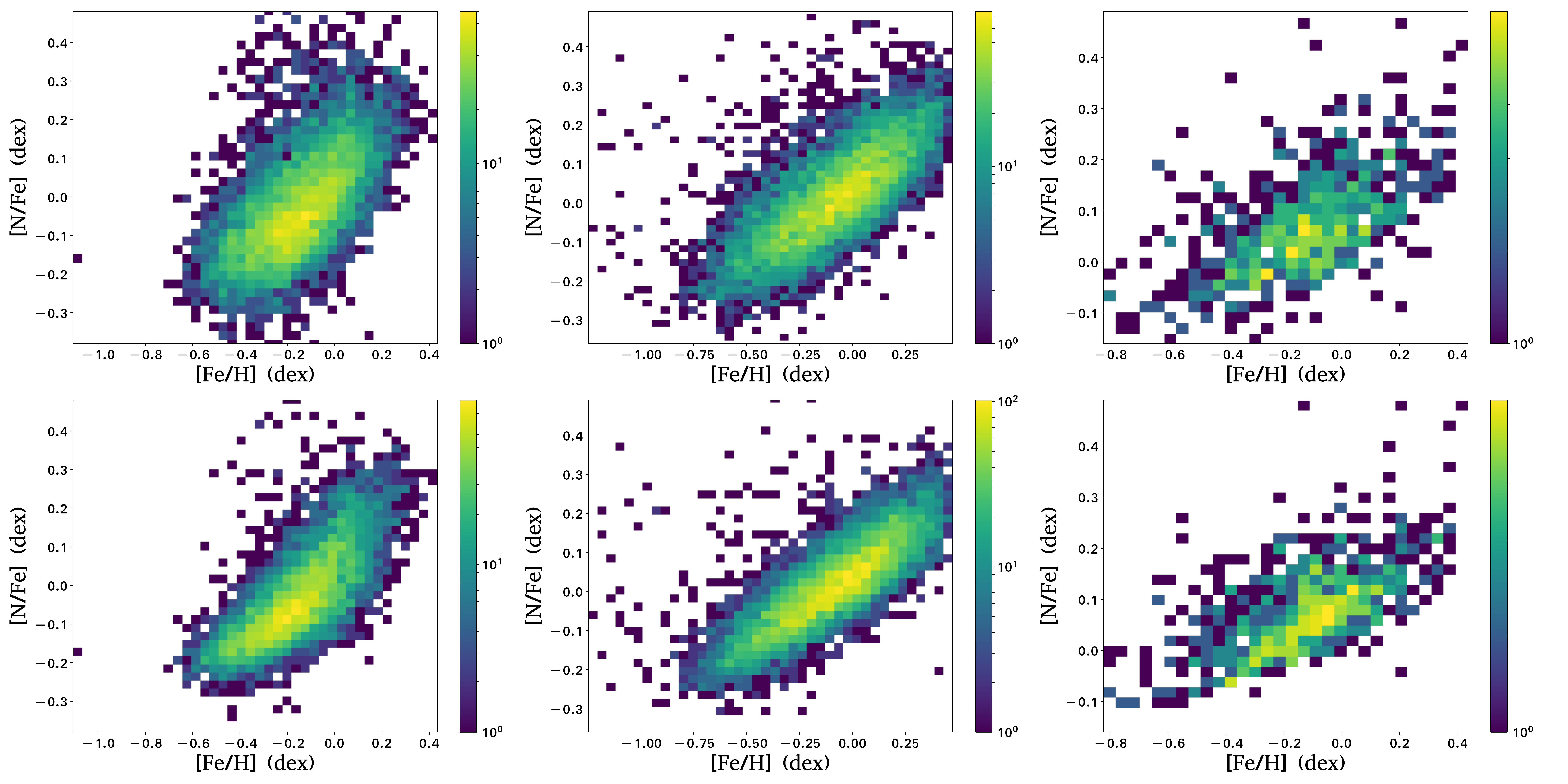}
  \caption{Similar to Fig.\,\ref{fig:nfe_giant}, but for the dwarf stars. The $BP-RP$ ranges are [0.50, 0.80] ({\it left} column), [0.80, 1.10] ({\it middle} column) and [1.10, 1.50] ({\it right} column), respectively.}
  \label{fig:nfe_dwarf}
\end{figure*}

\begin{figure*}
  \centering
  \includegraphics[width=\textwidth]{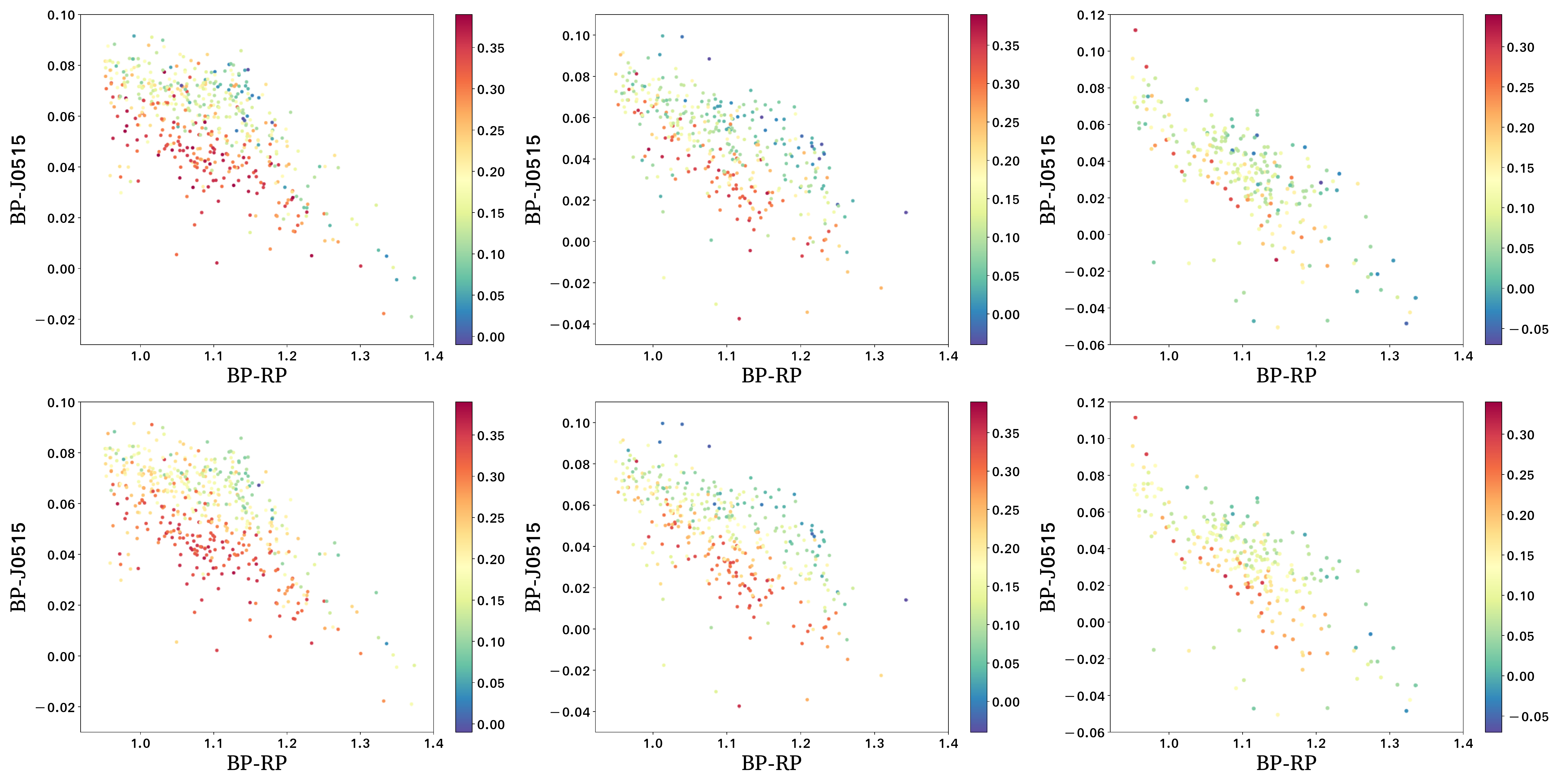}
  \caption{Distributions of [Mg/Fe] in the [$BP-RP$] -- [$BP-J0515$] color-color diagram from the LAMOST {\it DD--Payne} ({\it top} panels) and the {\it CSNet} results ({\it bottom} panels) for the training/testing sample giant stars, all color-coded by [Mg/Fe]. Different columns are for stars of different 
  [Fe/H] ranges. From left to right these are [$-$0.5, $-$0.3], [$-$0.3, $-$0.1], and [$-$0.1, 0.1], respectively.}
  \label{fig:mgfe_0515_giant}
\end{figure*}

\begin{figure*}
  \centering
  \includegraphics[width=\textwidth]{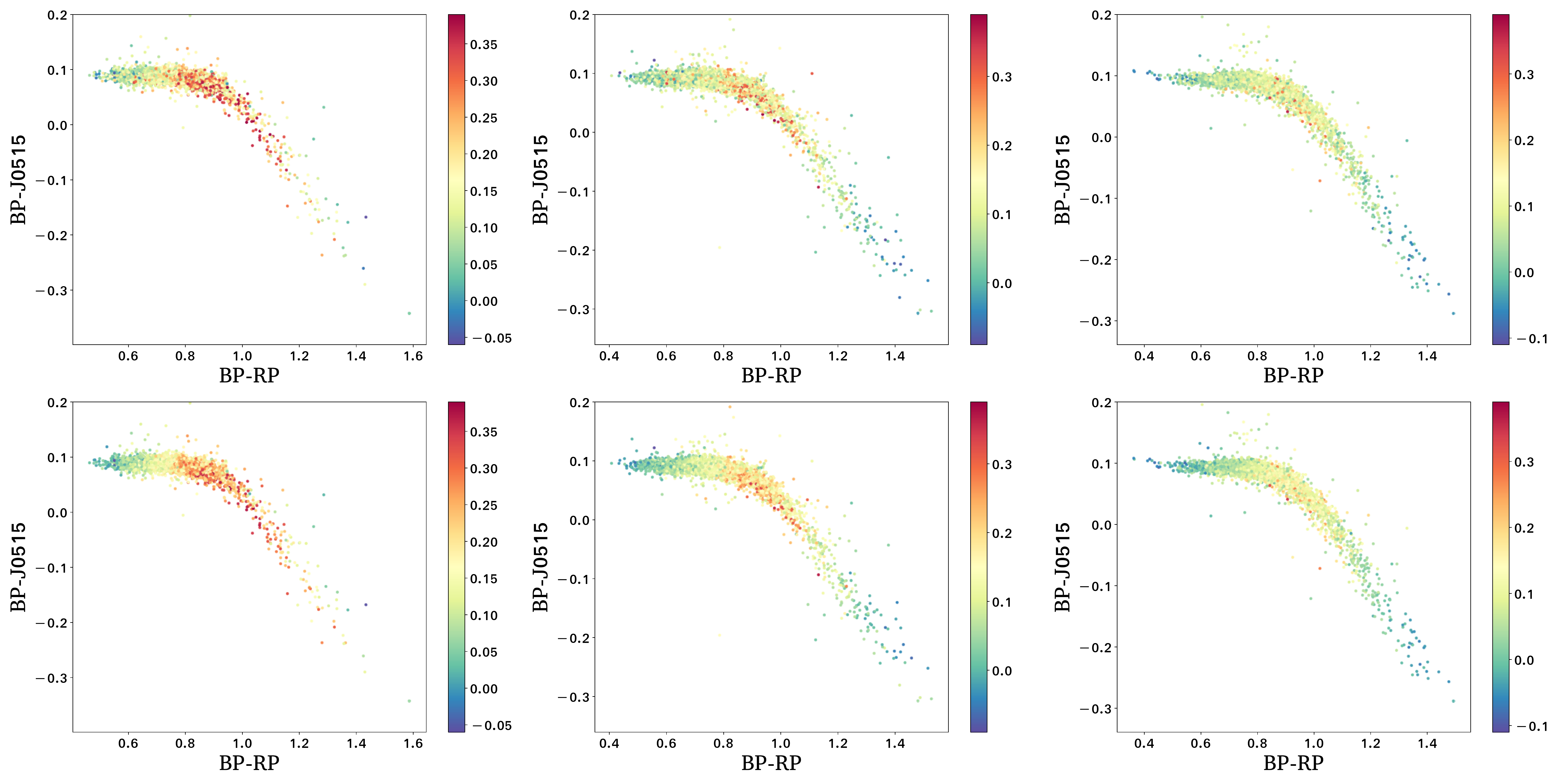}
  \caption{Similar to Fig.\,\ref{fig:mgfe_0515_giant}, but for the dwarf stars.}
  \label{fig:mgfe_0515_dwarf}
\end{figure*}

\begin{figure*}
  \centering
  \includegraphics[width=\textwidth]{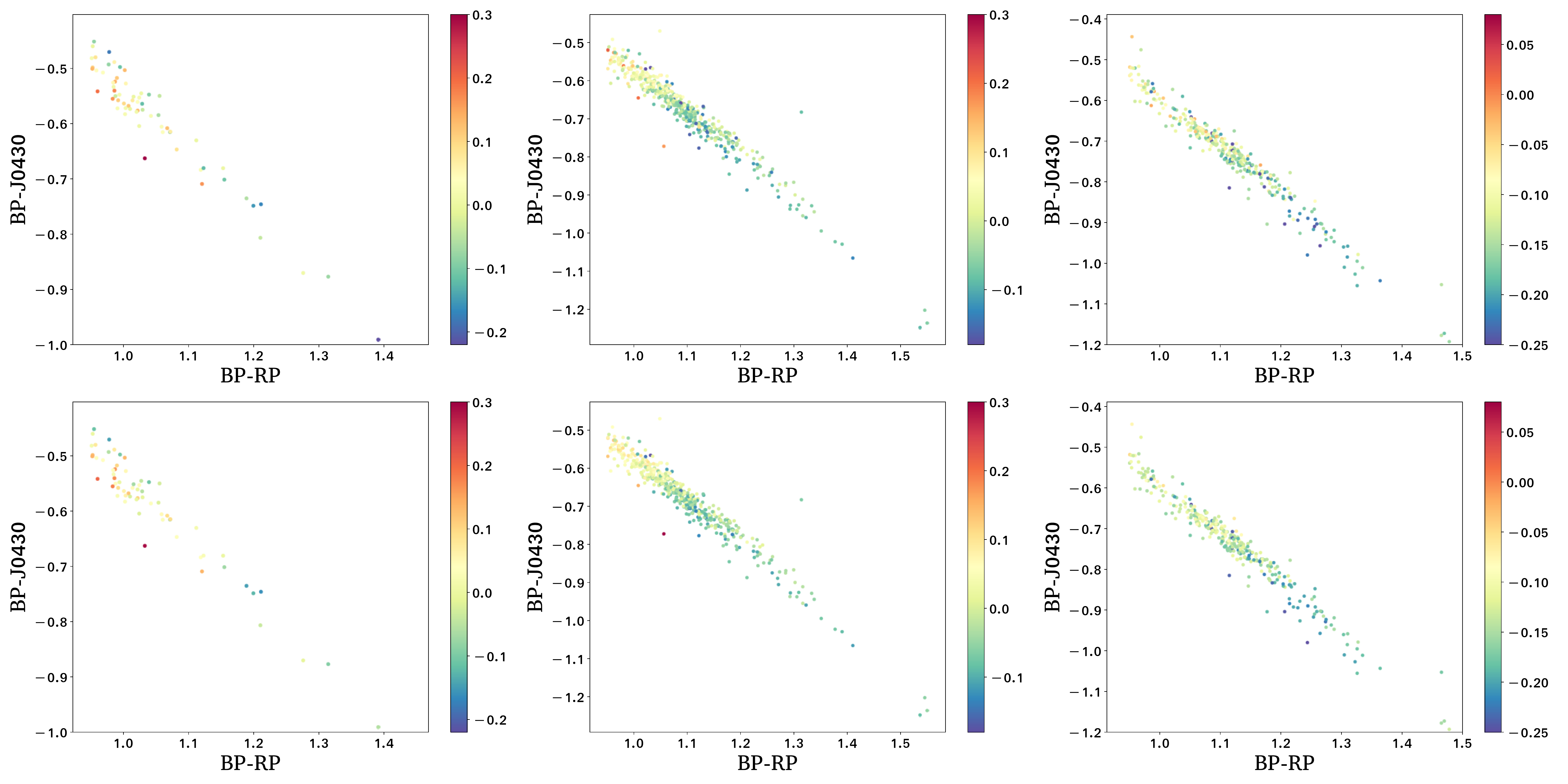}
  \caption{Distributions of [C/Fe] in the [$BP-RP$] -- [$BP-J0430$] color-color diagram from the LAMOST {\it DD--Payne} ({\it top} panels) and the {\it CSNet} results ({\it bottom} panels) for the training/testing sample giant stars, all color-coded by [C/Fe]. Different columns are for stars of different 
  [Fe/H] ranges. From left to right these are [$-$1.2, $-$0.9], [$-$0.7, $-$0.5], and [$-$0.1, 0.1], respectively.}
  \label{fig:cfe_0430_giant}
\end{figure*}

\begin{figure*}
  \centering
  \includegraphics[width=\textwidth]{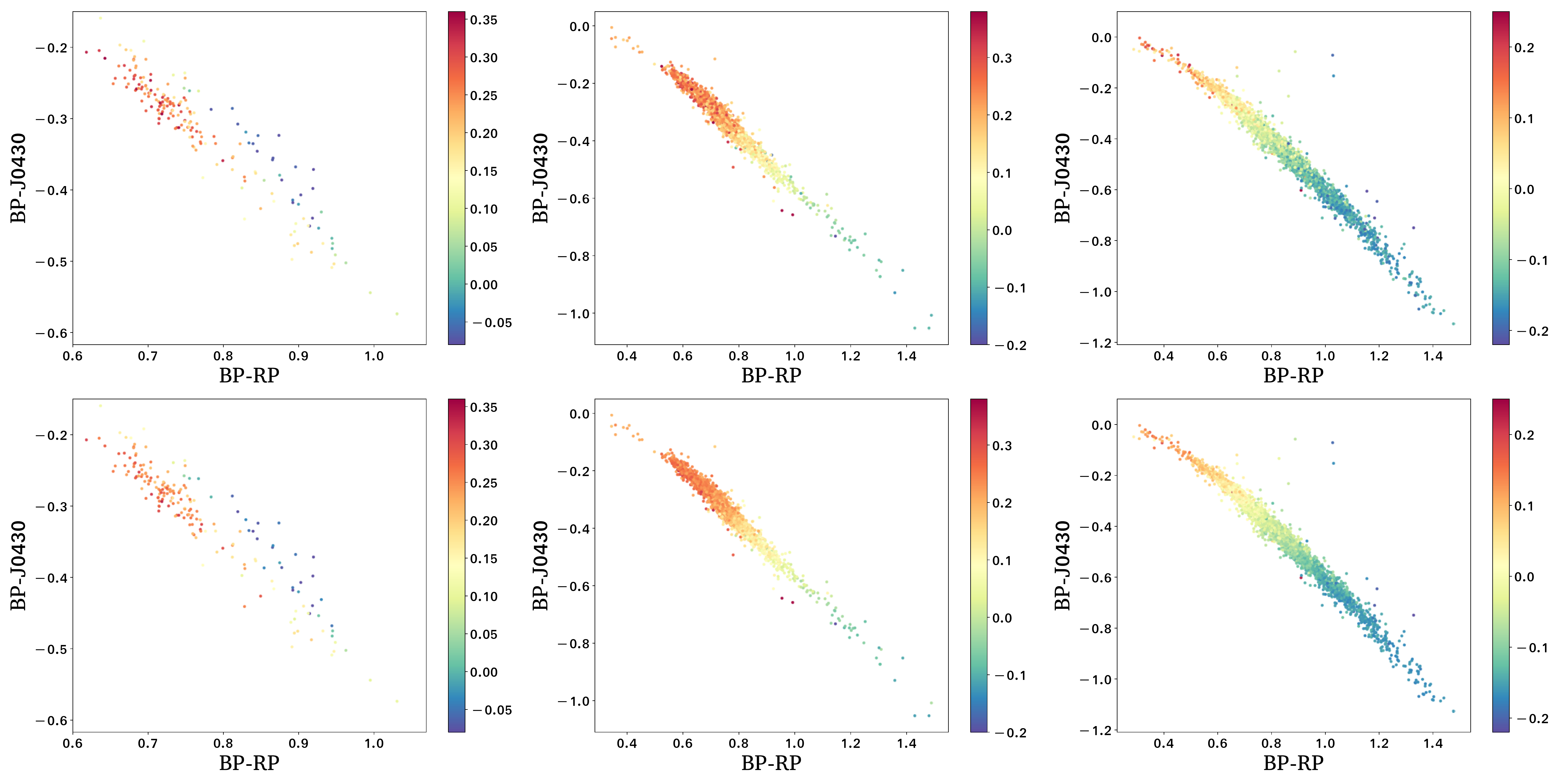}
  \caption{Similar to Fig.\,\ref{fig:cfe_0430_giant}, but for the dwarf stars.}
  \label{fig:cfe_0430_dwarf}
\end{figure*}

\begin{figure*}
  \centering
  \includegraphics[width=\textwidth]{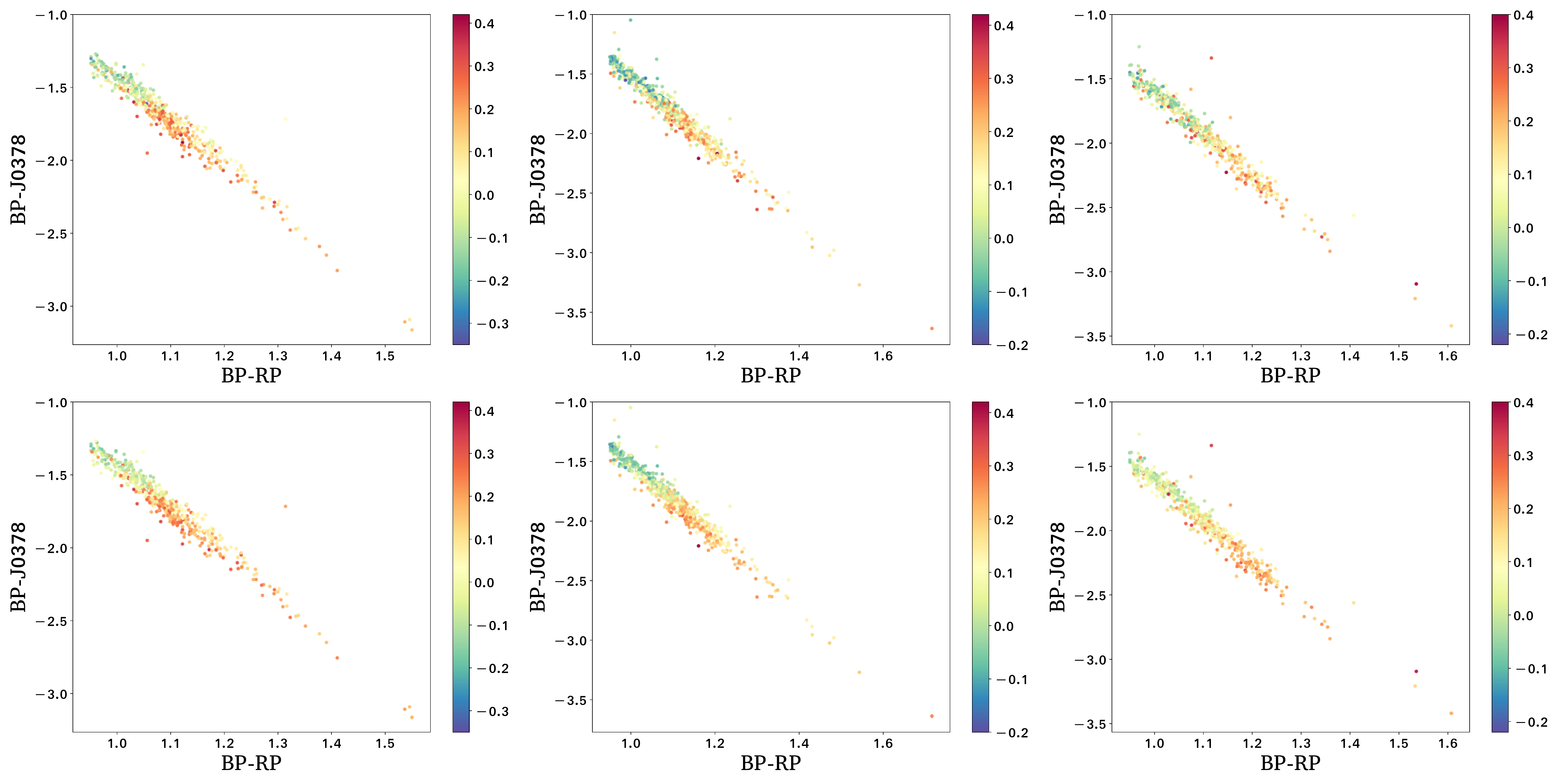}
  \caption{Distributions of [N/Fe] in the  [$BP-RP$] -- [$BP-J0378$] color-color diagram from the LAMOST {\it DD--Payne} ({\it top} panels) and the {\it CSNet} results ({\it bottom} panels) for the training/test giant stars, all color-coded by [N/Fe].  Different columns are for stars of different 
  [Fe/H] ranges. From left to right these are [$-$0.7, $-$0.5], [$-$0.5, $-$0.3], and [$-$0.3, $-$0.1], respectively.}
  \label{fig:nfe_0378_giant}
\end{figure*}

\begin{figure*}
  \centering
  \includegraphics[width=\textwidth]{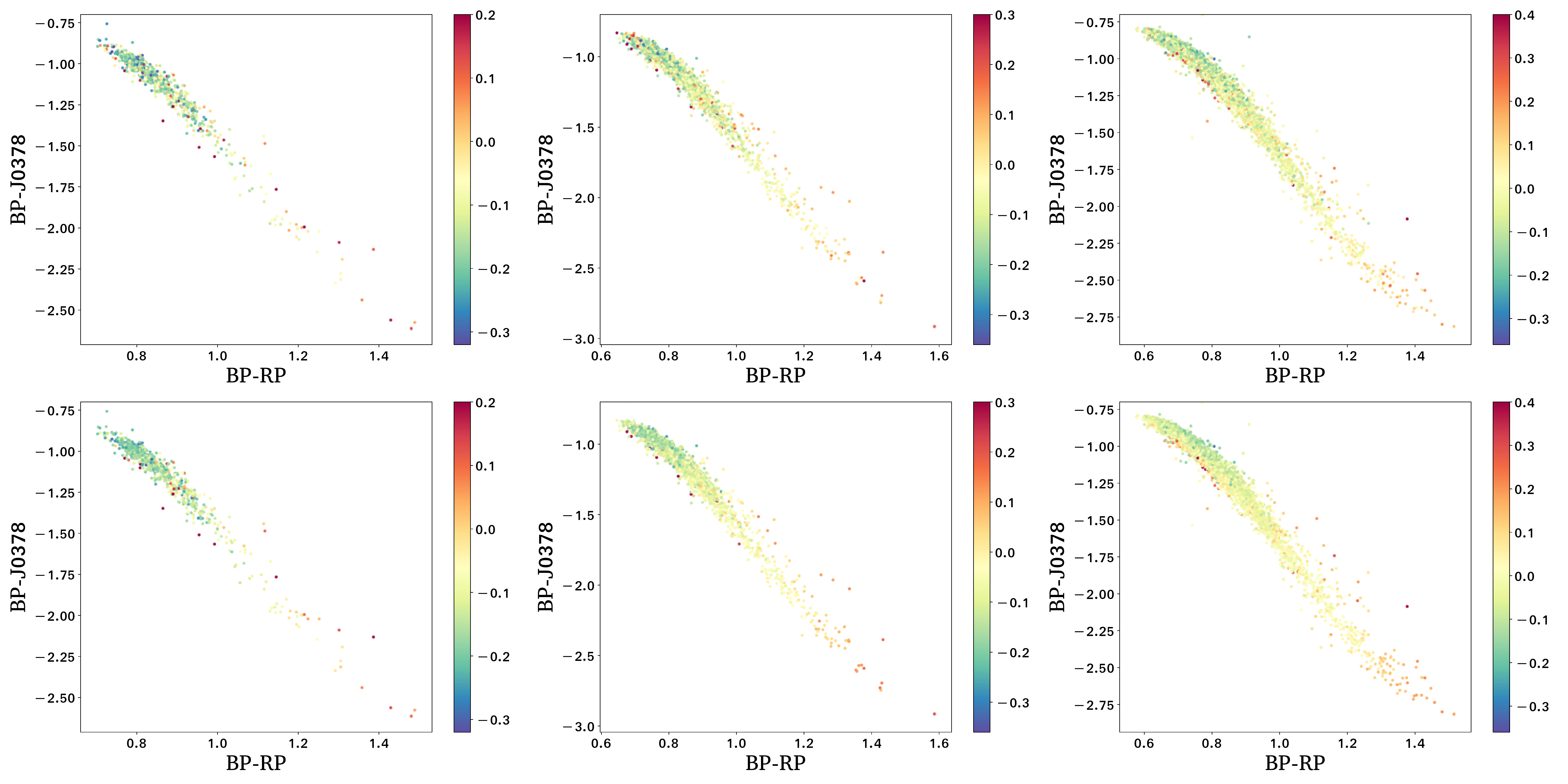}
  \caption{Similar to Fig.\,\ref{fig:nfe_0378_giant}, but for the dwarf stars.}
  \label{fig:nfe_0378_dwarf}
\end{figure*}

Figs.\,\ref{fig:mgfe_giant}--\ref{fig:nfe_dwarf} show comparisons for [Mg/Fe], [C/Fe], and [N/Fe] between
our results and LAMOST {\it DD--Payne} in the plane of [X-Fe]--[Fe/H]. The differences between 
the {\it CSNet} results and the  LAMOST {\it DD--Payne} are very small.
Figs.\,\ref{fig:mgfe_0515_giant}--\ref{fig:nfe_0378_dwarf} show comparisons in the color-color diagrams
($BP-J0515$ -- $BP-RP$ for [Mg/Fe], $BP-J0430$ -- $BP-RP$ for [C/Fe], and $BP-J0378$ -- $BP-RP$ for [N/Fe]). We can see that abundances of our results are all consistent with those of LAMOST {\it DD--Payne},  indicating that the training process of {\it CSNet} works as expected. 
At a given narrow [Fe/H] range and $BP-RP$ color,
correlations between $BP-J0515$ and [Mg/Fe], $BP-J0430$ and [C/Fe], $BP-J0378$ and [N/Fe] are significant, demonstrating that {\it CSNet} measures these elemental abundances from $ab\,initio$ features, rather than drawing on astrophysical correlations among the stellar labels.     

\label{sect:Relativity}

\section{Comparisons of LAMOST with APOGEE--{\it Payne}, APOGEE--{\it ASPCAP} and GALAH--{\it Cannon} stellar labels}\label{LAMOST DR5 errors}

\begin{figure*}
  \centering
  \includegraphics[width=\textwidth]{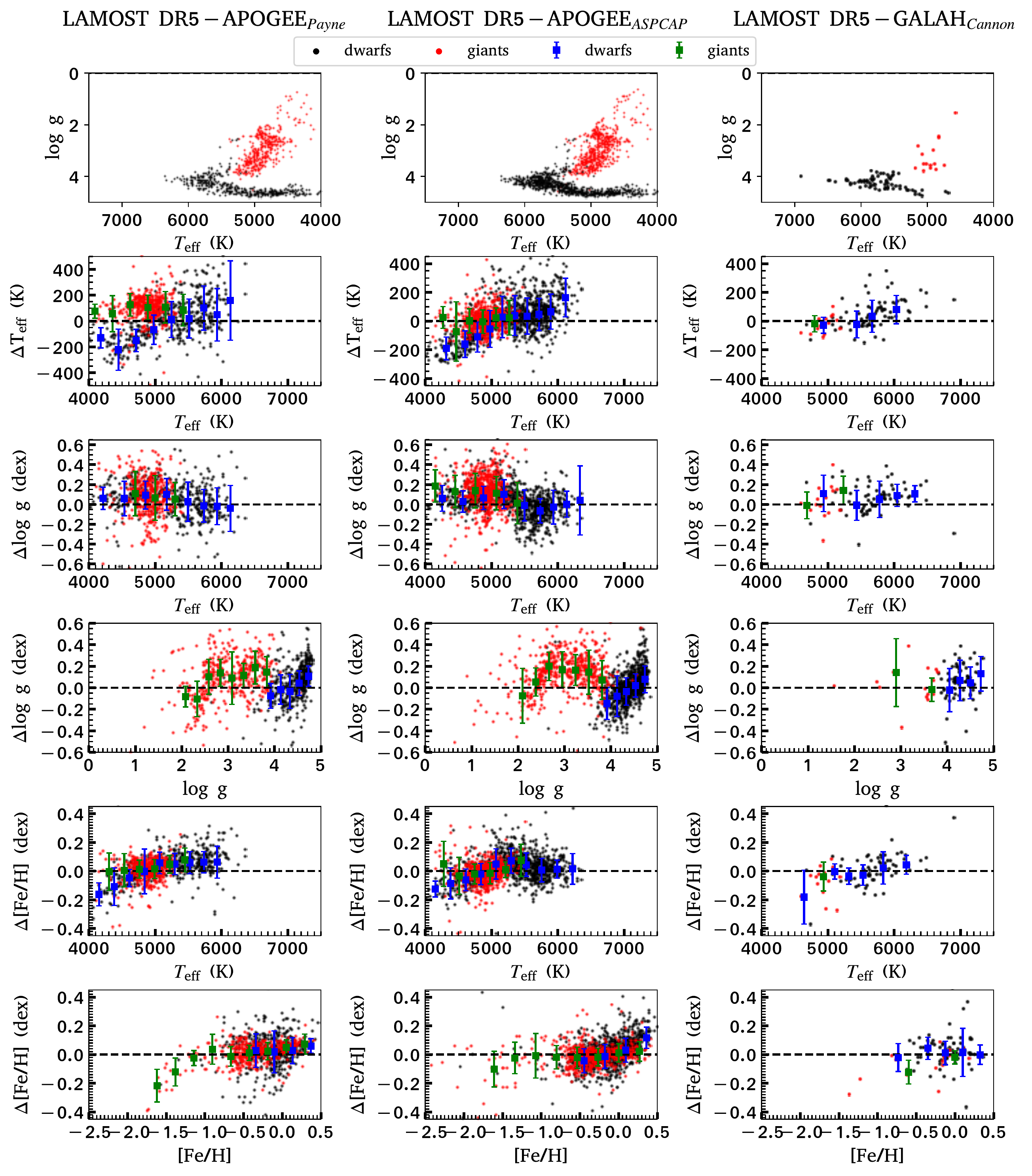}
  \caption{Comparisons for stellar atmospheric parameters including $T_{\rm eff}$, $\log g$, and [Fe/H]  between the LAMOST DR5 and the APOGEE--{\it payne} (left column), APOGEE--{\it ASPCAP} (middle column), and GALAH--{\it Cannon} (right column). Only stars in commn with the LAMOST training/test
  samples are used. The black and red dots represent dwarfs and giants, respectively. Error bars, colored by blue and green for dwarfs and giants respectively, and indicate the mean value ``bias” and 1 $\sigma$ uncertainty of the residuals estimated using Gaussian fit.}
  \label{fig:label_cross_parameter}
\end{figure*}

\begin{figure*}
  \centering
  \includegraphics[width=\textwidth]{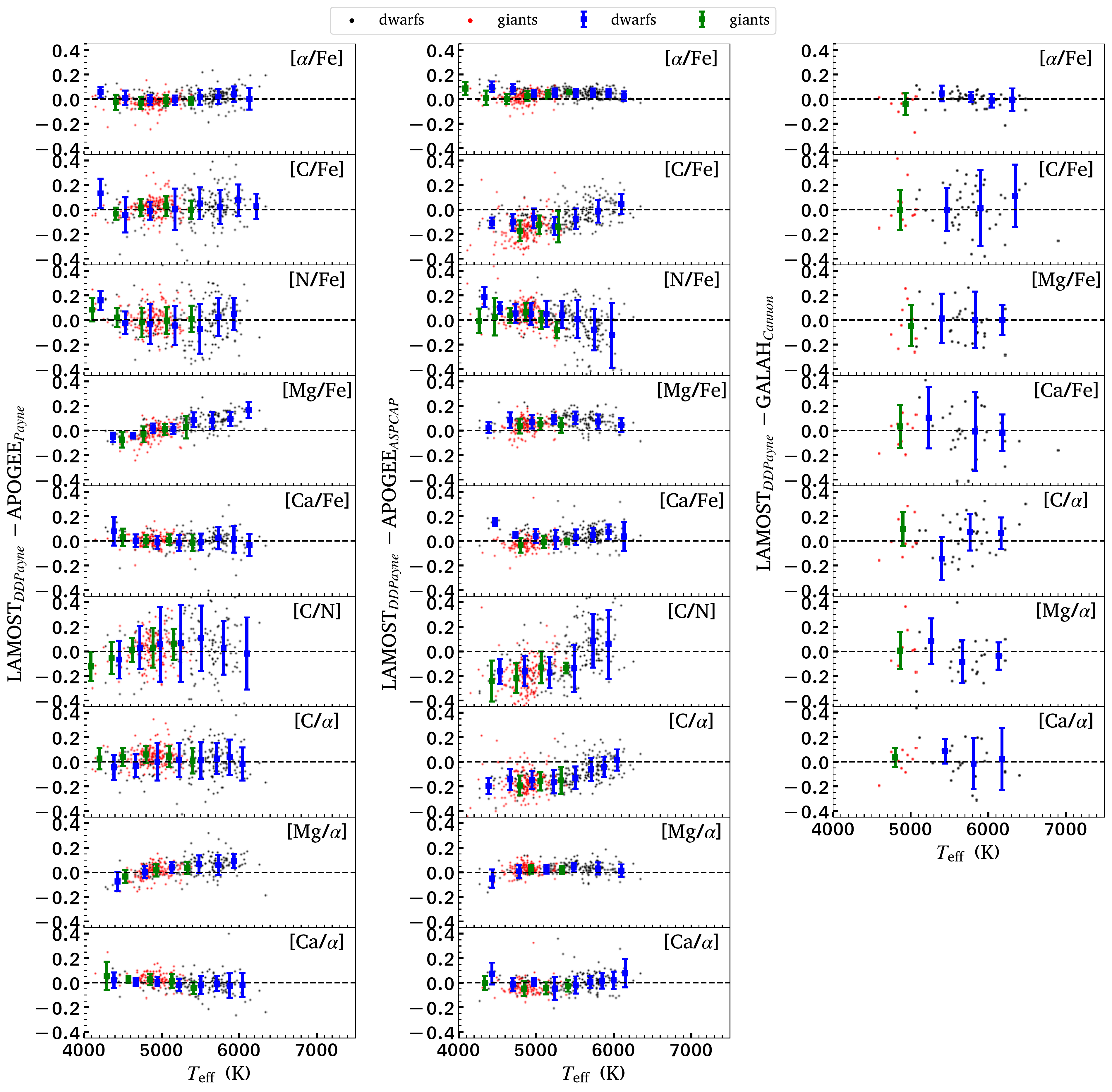}
  \caption{Similar to Fig.\,\ref{fig:label_cross_parameter}, but for the elemental abundances as a function of $T_{\rm eff}$. Note that here the LAMOST elemental abundances are from the DD--{\it Payne}.}
  \label{fig:label_cross_abundance}
\end{figure*}

We have shown in this study that systematic errors between {\it CSNet} and the validation samples are inherited from the training sets, rather than from the models themselves. Figs.\,\ref{fig:label_cross_parameter} and Figs.\,\ref{fig:label_cross_abundance} show comparisons of LAMOST catalogs and other surveys (APOGEE--{\it }, APOGEE--{\it ASPCAP} and GALAH--{\it Cannon}) for basic stellar atmospheric parameters and elemental abundances, respectively. Systematic discrepancies and trends for stellar labels between the LAMOST catalogs and above three surveys are found. More importantly, these patterns are consistent with those shown in the {\it CSNet} results (Figs.\,\ref{fig:cross_parameter} and Figs.\,\ref{fig:cross_abundance}), suggesting that the systematic offsets shown in the main text are inherited from the training data.   

\section{Stellar label distributions of LAMOST}\label{LAMOST DR5 distributions}

\begin{figure*}
  \centering
  \includegraphics[width=\textwidth]{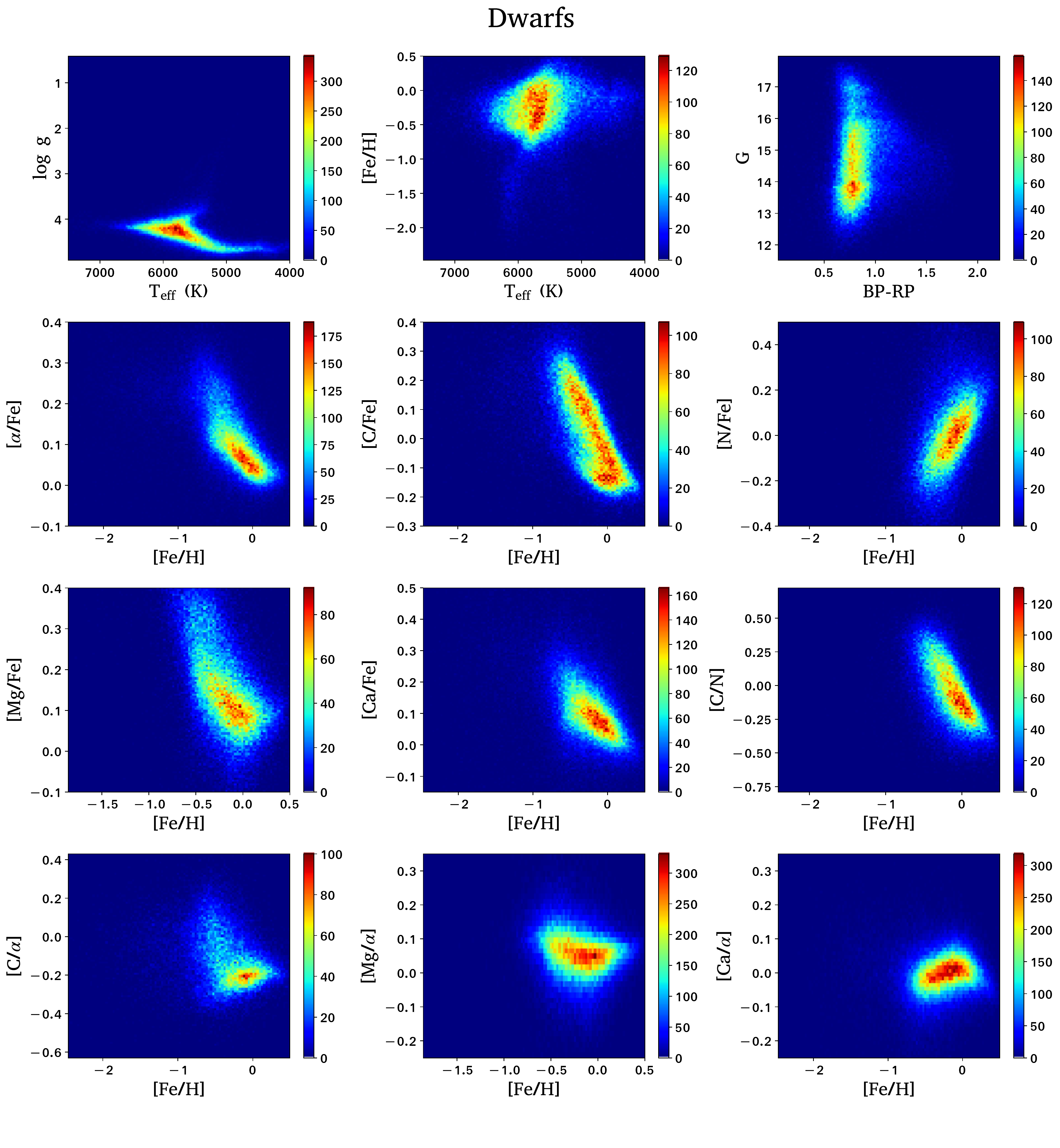}
  \caption{
  Density distributions of selected J-PLUS-{\it Gaia}-LAMOST dwarf stars in the planes of $T_{\rm eff}$--$\log g$, $T_{\rm eff}$--[Fe/H], ($BP-RP$)--$G$, and different elemental abundances with respect to [Fe/H], all color-coded by stellar number density. The stellar 
  labels are from LAMOST.
  Only stars that satisfy the following criteria are used: (1) FLAGS = 0; (2) 0.063 < BP$-$RP < 1.786; (3) $G<18$; (4) err (all J-PLUS filters) < 0.1 mag.
  }
  \label{fig:jplusgaia_d_train}
\end{figure*}

\begin{figure*}
  \centering
  \includegraphics[width=\textwidth]{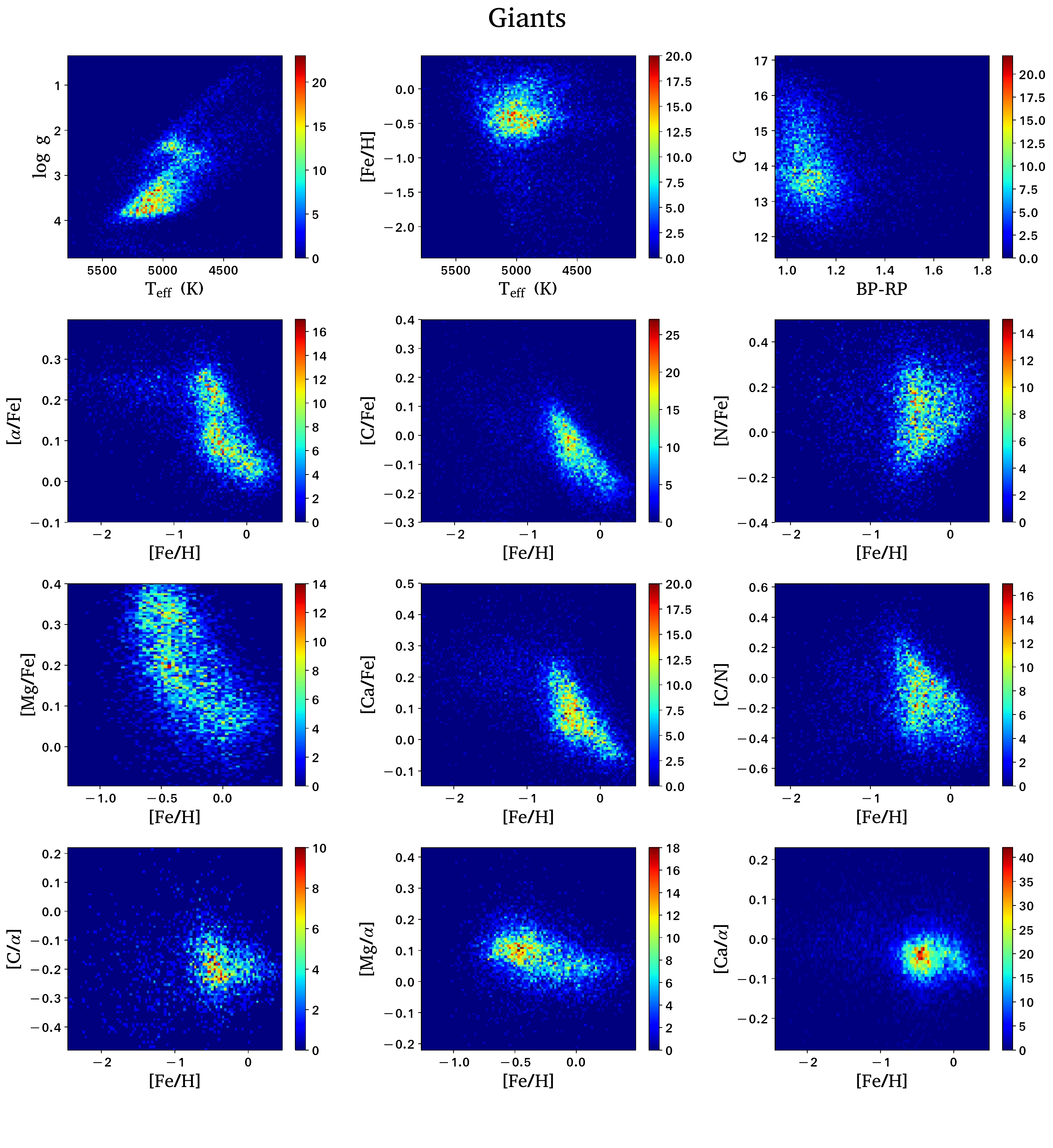}
  \caption{Similar to Fig.\,\ref{fig:jplusgaia_d_train}, but for giant stars.}
  \label{fig:jplusgaia_g_train}
\end{figure*}

Here we select stars in common between the J-PLUS DR1, {\it Gaia} DR2, and LAMOST DR5. For these stars 
we further apply the same criteria as those of Fig.\,\ref{fig:jplusgaia_d}: 
(1) FLAGS = 0; (2) $0.063 < BP-RP < 1.786$; (3) $G<18$; (4) err (all J-PLUS filters) < 0.1 mag.
Fig.\,\ref{fig:jplusgaia_d_train} and Fig.\,\ref{fig:jplusgaia_g_train} show the density distributions of the selected dwarfs and giants in the planes of $T_{\rm eff}$ -- $\log g$, $T_{\rm eff}$ -- [Fe/H], and ($BP-RP$) -- $G$, and different elemental abundances with respect to [Fe/H], with the stellar labels from LAMOST, respectively . The results are consistent with those shown in Fig.\,\ref{fig:jplusgaia_d} and Fig.\,\ref{fig:jplusgaia_g}. 

\end{appendix}

\end{document}